\documentclass[aps,prb,twocolumn,showpacs,superscriptaddress,10pt]{revtex4-1}
\usepackage{graphicx,amsmath,amssymb,amsfonts,sidecap,color}

\newcommand{\ket}[1]{\ensuremath{\left|#1\right\rangle}}

\makeatletter
\newcommand*{\rom}[1]{\expandafter\@slowromancap\romannumeral #1@}
\makeatother

\begin{document}

\title{Quantum Spin Hall Effect in Strip of Stripes Model}

\author{Jelena Klinovaja}
\affiliation{Department of Physics, Harvard University,  Cambridge, Massachusetts 02138, USA} 
\author{Yaroslav Tserkovnyak}
\affiliation{Department of Physics and Astronomy, University of California, Los Angeles, California 90095, USA}

\date{\today}
%\pacs{71.10.Pm; 74.45.+c; 05.30.Pr; 73.21.Hb}
%73.63.Nm	Quantum wires for transport
%74.45.+c	Proximity effects; Andreev reflection; SN and SNS junctions
%71.10.Fd 	Lattice fermion models (Hubbard model, etc.) 
% 71.10.Pm 	Fermions in reduced dimensions (anyons, composite fermions, Luttinger liquid, etc.) (for anyon mechanism in superconductors, see 74.20.Mn)
%73.21.Cd 	Superlattices 
%73.21.Hb 	Quantum wires
% 73.21.La 	Quantum dots 
%05.30.Pr	Fractional statistics systems (anyons, etc.
% 74.20.-z 	Theories and models of superconducting states

\begin{abstract}
We consider quantum spin Hall effect in an anisotropic strip of stripes and address both integer and fractional filling factors. The first model is based on a gradient of spin-orbit interaction in the direction perpendicular to the stripes. The second model is based on two weakly coupled strips with reversed dispersion relations. We demonstrate that these systems host helical modes, modes in which opposite spins propagate in  opposite directions. In the integer regime, the modes carry an elementary electron charge whereas in the fractional regime they carry fractional charges, and their excitations possess anyonic braiding statistics. These simple quasi-one-dimensional models can serve as a platform for understanding effects arising due to electron-electron correlations in topological insulators.
\end{abstract}

\maketitle

\section{Introduction}

Topological properties of condensed matter systems have been attracting attention already for several decades. Beginning with the fractional fermions in the Jackiw-Rebbi and the Su-Schrieffer-Heeger (SSH) models,\cite{Jackiw_Rebbi,FracCharge_Su,FracCharge_Kivelson,CDW,Two_field_Klinovaja_Stano_Loss_2012,Klinovaja_Loss_FF_1D,FF_transport} shifting  later to the quantum Hall effect\cite{Klitzing,Tsui_82,QHE_Review_Prange,book_Jain,Laughlin_FQHI,Halperin,Lebed,Montambaux,Lebed_Gorkov,Yakovenko_PRB,Lee_PRB,Yakovenko_review,Kane_PRL,Kane_PRB,Stripes_PRL,Stripes_arxiv,Stripe_PRL_exp} and topological insulators,
\cite{Hasan_review,Volkov_TI1,Volkov_TI2,Volkov_TI3,Fu_Kane,Zhang_TI,Patric_TI,Carlos_TI,Konig_2,Konig,Roth_TI,exp_1,Nowack_TI,Amir_TI,TI_Ady,Bernevig,Kane_Mele_graphene,Zhang_RMP,oreg,Neupert,Chamon_Mudry_2011,TI_inverted,Du_exp,Ady_wires} and finally getting enriched by the bound states with non-Abelian braiding statistics, Majorana fermions\cite{Read_2000,fu,Nagaosa_2009,Sato,demler_2011,potter_majoranas_2011,
lutchyn_majorana_wire_2010,oreg_majorana_wire_2010,alicea_majoranas_2010,
RKKY_Basel,RKKY_Simon,RKKY_Franz,Klinovaja_CNT,bilayer_MF_2012,MF_nanoribbon,MF_MOS,MF_Bena,MF_ee_Suhas,Ando,mourik_signatures_2012,deng_observation_2012,das_evidence_2012,Rokhinson,Goldhaber,marcus_MF}  and parafermions,\cite{Fradkin_PF_1980,topology_barkeshli,barkeshli_2, Fendley_PF_2012,PF_Linder,Cheng,Vaezi,PF_Clarke,
Ady_FMF,PF_Mong,vaezi_2,PFs_Loss,PFs_Loss_2,PF_TI_Amir,Vaezi_Barkeshli} the field continues its rapid development. Not only pure scientific curiosity but also the promise of practical applications, such as conventional and topological quantum computing, provides strong motivation to explore this topic.

 The theoretical description of many spectacular experiments, such as, for example, the precise quantization of the Hall conductance,\cite{Klitzing,Tsui_82}  relies on effective models. This is especially true in the case of two-dimensional systems with strong electron-electron interactions, such as fractional quantum Hall effect and fractional topological insulators.
 At the same time, the quantum Hall effect has been observed not only in the usual two-dimensional electron gases created in semiconductor heterostructures\cite{Klitzing,Tsui_82}  but also in quasi-two-dimensional materials such as graphene\cite{QHE_graphene} and organic compounds, for example, Bechgaard salts. \cite{Lebed,Montambaux,Lebed_Gorkov,Yakovenko_PRB,Lee_PRB,Yakovenko_review,Stripe_PRL_exp} The main difference of the latter materials, which could be turned to a conceptual advantage, is their conduction anisotropy, such that the effective mass (effective hopping matrix element) in the $x$ direction is much larger (smaller) than in the perpendicular $y$ direction. Hence, 
these materials can be treated as a strip of coupled stripes that allows one to treat tunneling between stripes as a small perturbation.\cite{Yakovenko_review} 
This representation of two-dimensional electron gas as a system of coupled one-dimensional channels or wires turned out to be very fruitful for including electron-electron interactions and, in particular, for constructing a description of the fractional quantum Hall effect in anisotropic systems.\cite{Kane_PRL,Kane_PRB,Stripes_PRL,Stripes_arxiv,Neupert,oreg} Moreover, this method is not specific for quantum Hall systems and can be generalized to other topological phases of two-dimensional systems. \cite{Neupert,oreg,Ady_wires} 
In particular, 
we implement these ideas to describe time-reversal invariant systems such as fractional topological insulators. 

In the present work,  we 
propose two systems which exhibit both integer and fractional quantum spin Hall effect, and, hopefully, can be realized in experiments using available ingredients. As a prototype of an effective one-dimensional channel, we consider nanowires, quantum wires created by gating, atomistic chains, and cold atom systems. In our work, we refer to these channels as stripes. This choice is determined by the fact that our prime focus is on one-dimensional channels created by gating inside two-dimensional systems or occurring spontaneously as density modulations due to surface reconstruction again caused by electron-electron interactions.

 The first model consists of a strip of stripes with Rashba spin orbit interaction (SOI), whose strength grows linearly from stripe to stripe. The direction of the corresponding SOI vectors that determines the spin polarization axis is the same for all stripes.
Such gradient of SOI can be created by a gradient in the electric field applied perpendicular to the strip.  The main advantage of this simple model lies in the fact that its topological properties can be easily seen by making a straightforward connection to the quantum Hall effect (QHE). In this sense, it serves here as a warm-up before we introduce  the second, much more complex and interesting model.
 Indeed, such non-uniform SOI is gauge-equivalent to the orbital effect produced by an effective magnetic field whose sign depends on the spin projection. The field is in the positive (negative) direction for spin up (down) states, so that we deal with two copies of quantum Hall effect. This brings us back in the analogy with the quantum spin Hall effect (QSHE) in graphene described by Kane and Mele\cite{Kane_Mele_graphene} as two copies of the Haldane model.\cite{Haldane} Also, an effective magnetic field allows us to introduce a concept of a filling factor $\nu$ for the QSHE by analogy with the QHE.
For example, at $\nu=1$ there is one pair of helical modes, modes which carry opposite spins in opposite directions. At the fractional filling factors such as $\nu=1/3$, elementary excitations in these modes possess fractional charges and non-trivial commutation relations. We note that, similarly to the fractional quantum Hall effect (FQHE), the fractional quantum spin Hall effect (FQSHE) is present only if electron-electron interactions are strong enough to generate dominant back-scattering terms.\cite{Kane_PRL}

The second system,  which is the main result of this work, is a bilayer composed of two strips that are coupled by tunneling.  The SOI inside a stripe is assumed to be uniform.  This uniformity is an important key property  that favours the bilayer model  over the non-uniform one as it opens up a new class of effective models that captures the essential features of  two-dimensional topological insulators currently observed in experiments.
 Moreover, in contrast to the previous model,  there is also SOI associated with propagation in the direction perpendicular to the stripes, such that interstrip tunneling is not spin-conserving. We show that an interplay between inter- and intrastrip tunneling results in the QSHE phase, if the intrastrip tunneling is dominant. Again, in the regime of strong electron-electron interactions, the system could exhibit FQSHE.

\section{Quantum Spin Hall Effect based on gradient of spin orbit interaction}

\subsection{Model}

\begin{figure}[!b]
\includegraphics[width=0.9\linewidth]{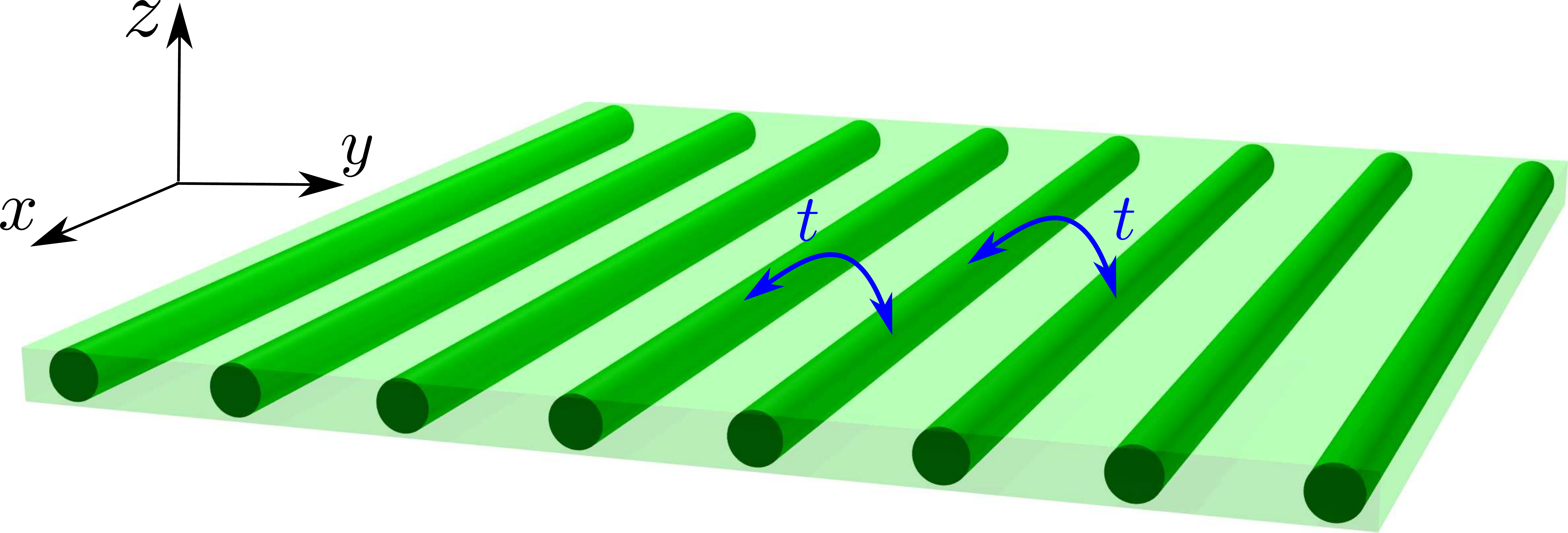}
\caption{The model consists of a strip of stripes stacked in the $y$ direction whereas the stripes stretch in the $x$ direction. 
Tunneling of the strength $t$ couples  neighbouring stripes.  The stripes, which can be considered as density modulations, could be formed inside two-dimensional structures.} 
\label{model}
\end{figure}

We consider a strip of stripes that is aligned in the $y$ direction 
and consists of a number of stripes (effective one-dimensional channels: also could be thought of as wires) that are aligned in the $x$ direction,\cite{Stripes_PRL,Stripes_arxiv}
 see Fig.~\ref{model}. (We imagine a generic system where the stripes might be embedded in some host material that allows tunneling between the stripes
and thus prefer the term stripe over the term wire).
We also allow for tunneling between two neighboring stripes. 
However, to simplify our calculations, we assume that the hopping matrix element $t$ between two neighboring stripes is small in comparison to the Fermi energy inside a stripe.
As a consequence, we first treat each stripe as completely isolated and only afterwards add the tunneling terms as a small perturbation.

The kinetic part of the Hamiltonian corresponding to the $n$th stripe is written as
\begin{align}
H_{0,n}= \sum_{\sigma=\pm1} \int dx\ \Psi_{n\sigma}^\dagger (x) \left(-\frac{\hbar^2\partial_x^2}{2m} -\mu\right)\Psi_{n\sigma} (x),
\end{align}
where the annihilation operator $\Psi_{n\sigma} (x)$ removes an electron (of charge $e$ and effective mass $m$)  with spin $\sigma=\pm 1$  at the point $x$ of the $n$th stripe. Here, we choose the spin quantization axis along the $z$ direction defined by the spin orbit interaction that is also present inside the stripes (see below).  The chemical potential $\mu$ is  uniform over the entire strip.

The Rashba spin orbit interaction (SOI) term 
acts inside each of the stripes and is written as
\begin{align}
H_{SOI,n} = -i \sum_{\sigma,\sigma'=\pm 1}\int dx\ \alpha_n \Psi_{n\sigma}^\dagger (x) (\sigma_3)_{\sigma\sigma'} \partial_x \Psi_{n\sigma'} (x),
\label{SOI}
\end{align}
where $\alpha_n$ is the SOI parameter that characterizes the SOI strength inside the $n$th stripe. The Pauli matrix $\sigma_i$ acts on spin space. In what follows we consider a strip with a gradient of spin-orbit interaction in the $y$ direction
\begin{align}
\alpha_n = (2n+1) \alpha_0.
\label{grad}
\end{align}
Such a gradient can be generated by a gradient in the electric field that induces the Rashba SOI.

The aforementioned tunneling between two neighboring stripes is described by
\begin{align}
H_{t} = t \sum_{n,\sigma=\pm1} \int dx\   \Psi_{(n+1)\sigma}^\dagger (x)  \Psi_{n\sigma} (x) + H.c.,
\label{y_tun}
\end{align}
where $t$ is a spin-conserving tunneling matrix element.

It is important to note that the SOI can be gauged away in strictly one-dimensional systems.\cite{RSOI_1,RSOI_2,braunecker_prb_10}  For example, by applying a spin-dependent gauge transformation defined for each wire separately as
\begin{equation}
\bar \Psi_{n\sigma} (x) = e^{i (2n+1) \sigma k_{so} x} \Psi_{n\sigma} (x),
\end{equation}
to the stripe Hamiltonian $H_{0,n}+H_{SOI,n}$, we absorb the SOI  into the kinetic term,
\begin{align}
\bar H_{0,n}= \sum_{\sigma=\pm1} \int dx\ \bar\Psi_{n\sigma}^\dagger (x) \left(-\frac{\hbar^2\partial_x^2}{2m} -\mu\right)\bar\Psi_{n\sigma} (x),
\end{align}
 where, assuming charge neutrality, we ignore the constant shift in energy. 
Here,  $k_{so} = m \alpha_0   /\hbar^2$  is the SOI wavevector of the stripe with  index $n=0$.
However, the tunneling term, which connects two neighboring stripes and makes the system effectively two-dimensional, keeps the information about the initial SOI. After the transformation, we arrive at 
\begin{align}
\bar H_{t} = t \sum_{n,\sigma} \int dx\  e^{2 i \sigma k_{so}x} \bar \Psi_{(n+1)\sigma}^\dagger (x)  \bar \Psi_{n\sigma} (x) + H.c.
\end{align}
where the tunneling matrix element $t$ acquires a phase and becomes spin- and position-dependent, i.e. $ t \to t_\sigma(x)= t e^{2i\sigma k_{so} x}$.  

Above we started with a model containing the gradient of SOI strength in the $y$ direction. However, as we have just seen, this model is equivalent to a model  with a special form of the SOI associated with the motion along the $y$ axis that results in a spin and position dependent phase $e^{2i\sigma k_{so} x}$ in the tunneling matrix element. We also note that the latter model can be understood in terms of orbital effects caused by two opposite magnetic fields ${\bf B}_\sigma = \sigma B \hat z$ that act on states with spin $\sigma=\pm 1$. Indeed, if the corresponding vector potential ${\bf A}_\sigma$, ${\bf B}_\sigma={\bf \nabla}  \times {\bf A}_\sigma$, is chosen to be in the $y$ direction, ${\bf A}_\sigma = \sigma Bx \hat y$, the phase acquired during the hopping from the $n$th stripe to the $(n+1)$th stripe is equal to $\phi_\sigma=(e/\hbar c)\int d{\bf r}\cdot  {\bf A}_\sigma \equiv 2  \sigma k_{so} x$. Here, we have chosen the strength of the magnetic field to be $B= 2k_{so}\hbar c/e a_y$,
 where $a_y$ is the distance between 
stripes. All these agrees with the SOI being treated as effective opposite orbital magnetic fields seen by electrons with opposite spins. Consequently, we arrive at two uncoupled coexisiting 2DEGs of spin-up and spin-down electrons in the quantum Hall regime similarly as it was done in the pioneering work of Kane and Mele \cite{Kane_Mele_graphene} based on the Haldane model. \cite{Haldane}  Thus, it is natural to expect that our simple model should also exhibit the quantum spin Hall effect. We confirm this  in the next two subsection and use this model as a conceptual warm-up before addressing more involved model introduced in the next section.

We also note that such an analogy of SOI with effective magnetic fields, allows us to introduce a concept of the filling factor $\nu$, as a ratio of the total spin up (spin down) electron number to the degeneracy of the Landau level in an effective magnetic field $B_\sigma$.

\subsection{Integer Quantum Spin Hall Effect}

\begin{figure}[!t]
\includegraphics[width=0.9\linewidth]{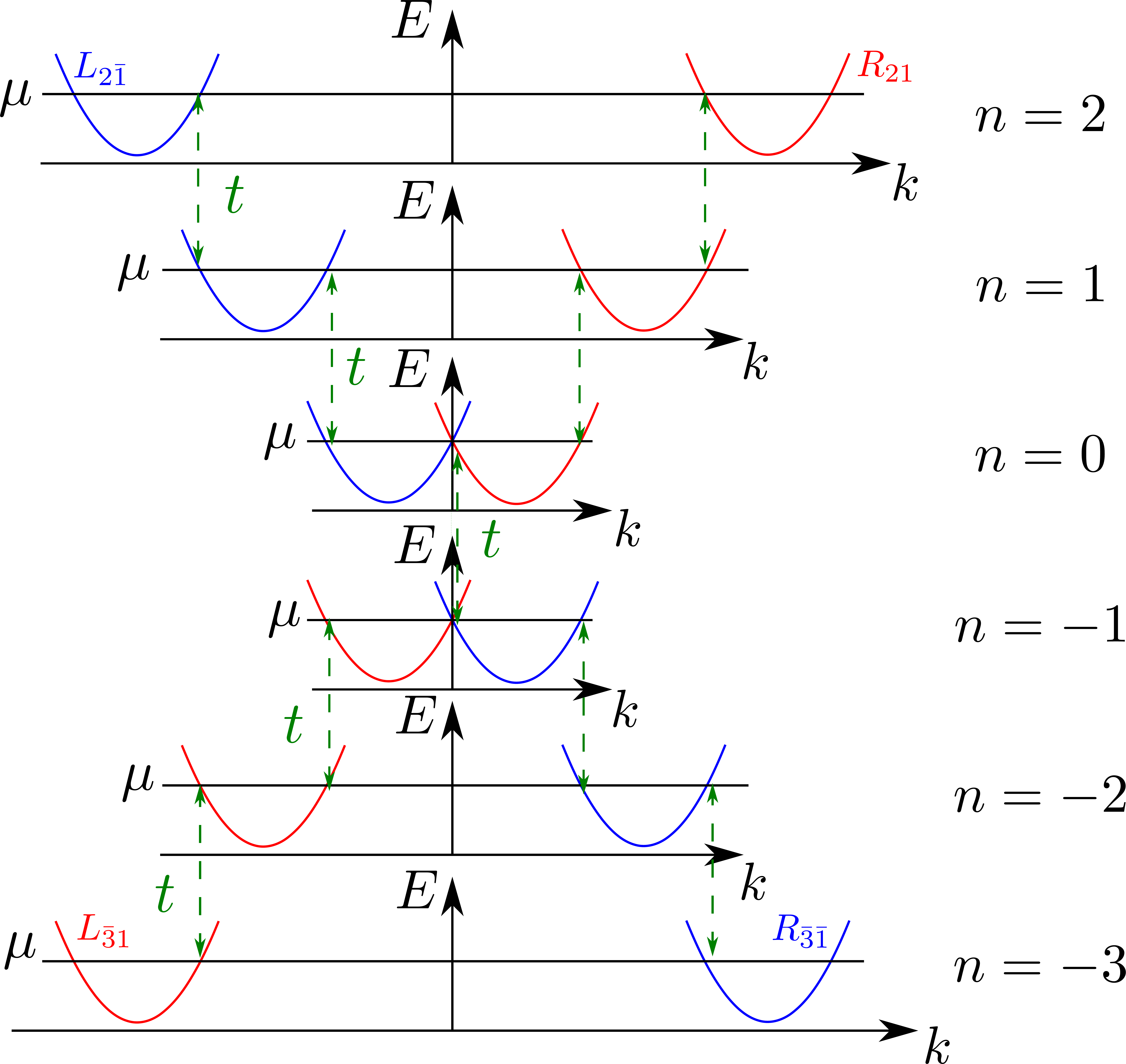}
\caption{The spectrum of a strip consisting of 6 stripes in the integer quantum spin  Hall regime characterized by the filling factor $\nu=1$. The chemical potential $\mu$ is uniform in the stripe and is tuned to the SOI energy of the zeroth stripe, $\mu=\mu_1$. The gradient of the Rashba SOI leads to a shift of Fermi wavevectors $k_{Fn\pm}$ from stripe to stripe such that $k_{Fn+}=k_{F(n+1)-}$. As a result, the tunneling $t$ between stripes is in resonance and couples right $R_{n \sigma}$ and left $L_{(n+\sigma)\sigma}$ movers. Consequently, the system is gapped in the bulk. However, there are two pairs of uncoupled modes left in the upper ($R_{21}$ and $L_{2\bar1}$) and lower ($L_{\bar3 1}$ and $R_{\bar 3\bar1}$) edges. These edge modes carry opposite spins in  opposite directions, which corresponds to the definition of the QSHE.\cite{Kane_Mele_graphene,Hasan_review}}
\label{ISHE1}
\end{figure}

We begin by assuming that  the chemical potential is tuned to the crossing point between the spin-up and spin-down branches of the stripe with index $n=0$, such that  $\mu = \mu_1\equiv  m \alpha_0^2/2\hbar^2$ is equal to the SOI energy, see Fig. \ref{ISHE1}. As a result, the Fermi wavevectors are given by $k_{Fn\pm} = k_{so} (2n +1 \pm 1)$, where $n$ is the stripe index.

{\it Edge modes in  $y$ direction.}
 Next, we focus on edge modes that propagate along the $y$ direction and are localized in the $x$ direction. However, for this we first study properties of the bulk energy spectrum.
For a moment, we assume that the system is periodic in the $y$ direction and contains $N_y$ stripes, so we can work with the Fourier transform characterized by the momentum $k_y$,
\begin{align}
 \bar\Psi_{k_y\sigma}(x) = \frac{1}{\sqrt{N_y}}\sum_{n} e^{ink_y a_y} \bar \Psi_{n\sigma}(x).
\end{align}
Hence, the total Hamiltonian is diagonal in momentum space
$H=\sum_{k_y} H_{k_y}$,
where $H_{k_y}=H_{0,k_y}+H_{t,k_y}$. Here, the kinetic term 
 is given by
\begin{align}
\bar H_{0,k_y}= \sum_{\sigma=\pm1} \int dx\ \bar\Psi_{k_y\sigma}^\dagger (x) \left(-\frac{\hbar^2\partial_x^2}{2m} -\mu\right)\bar\Psi_{k_y\sigma} (x),
\end{align}
and the tunneling term by
\begin{align}
&\bar H_{t,k_y} = t \sum_{\sigma} \int dx\  e^{2 i \sigma k_{so}x + ik_y a_y} \bar\Psi_{k_y\sigma}^\dagger (x)   \bar\Psi_{k_y\sigma} (x) \nonumber\\
&\hspace{170pt}+ H.c.
\end{align}

Close to the Fermi points $\pm k_{so}$, the operator $\bar\Psi_{k_y\sigma}(x)$ can be represented in terms of slowly-varying right-mover $R_{k_y,\sigma}(x)$  and left-mover $L_{k_y,\sigma}(x)$  fields as
\begin{align}
&\bar\Psi_{k_y\sigma}(x) = R_{k_y\sigma}(x) e^{i k_{so}x} + L_{k_y\sigma}(x) e^{-i k_{so}x},
\end{align}
 The kinetic term can then be rewritten as
\begin{align}
& H_{0,k_y}=-i\hbar \upsilon_F \sum_{\sigma=\pm1} \int dx\  \Big[R^\dagger_{k_y\sigma}(x) \partial_x   R_{k_y\sigma}(x) \\
&\hspace{130pt} -  L^\dagger_{k_y\sigma}(x) \partial_x   L_{k_y\sigma}(x) \Big] \nonumber,
\end{align}
where $\upsilon_F=\alpha_0/\hbar$ is the Fermi velocity.

The tunneling term $H_{t,k_y}$ becomes in this right-/left-mover representation
\begin{align}
&H_{t,k_y} = t \int dx\  \Big[e^{ ik_y a_y} R_{k_y1}^\dagger (x)  L_{k_y1} (x)\nonumber\\
&\hspace{60pt}+e^{ik_y a_y} L_{k_y\bar 1}^\dagger (x) R_{k_y\bar 1} (x) +H.c.\Big],
\end{align}
where we drop all fast oscillating terms.  We note here that without loss of generality we have neglected a possible SOI related to the motion in the $y$ direction [see Eq. (\ref{y_tun})], because even if present it would not change $H_{t,k_y}$ in the leading order. Moreover, it cannot lead to gaps in the spectrum on its own and, thus, does not play a crucial role for the present purpose.

The total Hamiltonian can be expressed in terms of the associated Hamiltonian density, $H_{k_y}=\int dx\ \Psi_{k_y}^\dagger(x) {\cal H}_{k_y} \Psi_{k_y}(x)$, where
\begin{align}
{\cal H}_{k_y} = \hbar \upsilon_F \hat k \lambda_3 + t [\cos (k_y a_y) \lambda_1-\sin (k_y a_y) \lambda_2 \sigma_3].
\label{ham_1}
\end{align}
Here, we choose the basis $\Psi_{k_y}(x)=(R_{k_y1}(x),L_{k_y1}(x),R_{k_y\bar 1}(x),L_{k_y\bar 1}(x))$ composed of the right- and left-movers. The momentum operator $\hat k= -i \partial_x$ is determined close to the Fermi points $\pm k_{so}$. The Pauli matrix $\lambda_i$ ($\sigma_i$) acts on right/left-mover (spin) space with $i=1,2,3$.

As we have already noted above, the bulk spectrum is fully gapped,
\begin{align}
E_{\pm} = \pm \sqrt{(\hbar \upsilon_F k)^2 + t^2},
\end{align}
where each energy level is twofold degenerate in spin. This degeneracy is nothing else but the Kramers degeneracy of energy levels in a time-reversal invariant system. In addition, the Hamiltonian ${\cal H}_{k_y}$ is block-diagonal in spin space, so spin $\sigma$ is a good quantum number, and eigenstates could be presented as spin-polarized. Moreover, ${\cal H}_{k_y}(\sigma) = {\cal H}_{-k_y}(-\sigma)$, and thus the spectrum and eigenstates for spin up can be obtained from the ones for spin down if one substitutes $k_y$ with $-k_y$. 

Next, we impose vanishing boundary conditions on the wavefunctions at the left and right ends of each stripe. For example, wavefunctions should go to zero at the left end of each stripe $x=0$: $\Phi(x=0)=0$. 
Using a standard scattering problem procedure of matching decaying eigenstate wavefunctions,  we arrive at the spectrum of edge modes $E_{k_y, \sigma}$,
\begin{align}
&E_{k_y, 1}= - t \cos (k_y a_y),\ \ \ k_y a_y \in (0, \pi),\\
&E_{k_y, \bar 1}= - t \cos (k_y a_y),\ \ \ k_y a_y \in (-\pi, 0).
\end{align}
As expected, all spin up (down) edge modes propagate with a positive (negative) velocity in the $y$ direction. The corresponding wavefunctions are given by
\begin{align}
&\Phi_{k_y,\sigma}(x)=\ket{\sigma} \sin (k_Fx) e^{-x/\xi_\sigma}, 
\end{align}
where $\xi_\sigma$ is the localization length, $\xi_\sigma = \sigma \hbar\upsilon_F/[t \sin (k_y a_y)]$.

As a result,  we constructed edge modes and explicitly confirmed that the system under consideration corresponds to the spin Hall system in which modes localized at the edges propagate in the direction determined by their spin.\cite{Kane_Mele_graphene,Hasan_review} In particular, in the constructed setup, the spin up mode possesses a positive velocity, whereas the spin down mode possesses a negative velocity along the left strip edge. Thus, such a system can be identified with QSHE at the filling factor $\nu=1$.

Next, we consider the QSHE for higher filling factors $\nu=m$ with $m$ being a positive integer.  In what follows we fix  the electron density and change the strength of SOI. Of course, alternatively, one can fix the SOI strength and tune the chemical potential. 
To achieve the regime of $\nu=m$, the gradient of the SOI (and, consequently, the corresponding effective magnetic field) should be $m$ times smaller than for the previously considered case $\nu=1$ [see Eq. (\ref{grad})],
\begin{align}
\alpha_n^{(m)} = \frac{(2n+1) \alpha_0}{m}.
\end{align} 
As a consequence, the gap opens only in  $m$th order of perturbation theory\cite{Stripes_arxiv} 
with  the size of the gap given by $t (t/E_F)^{(m-1)}$, where $E_F$ is the Fermi energy of the strip. Importantly, there are now $m$ spin up (down) modes\cite{Stripes_arxiv} propagating with positive (negative) velocities.   However, we should note that the system can still develop a full gap in the edge mode spectrum if there are an even number of helical pairs, and local perturbation terms allow for scattering between spin up and spin down states.\cite{Hasan_review} This brings us back to the $\mathbb Z_2$ classification.\cite{Neupert}

\begin{figure}[!t]
\includegraphics[width=0.9\linewidth]{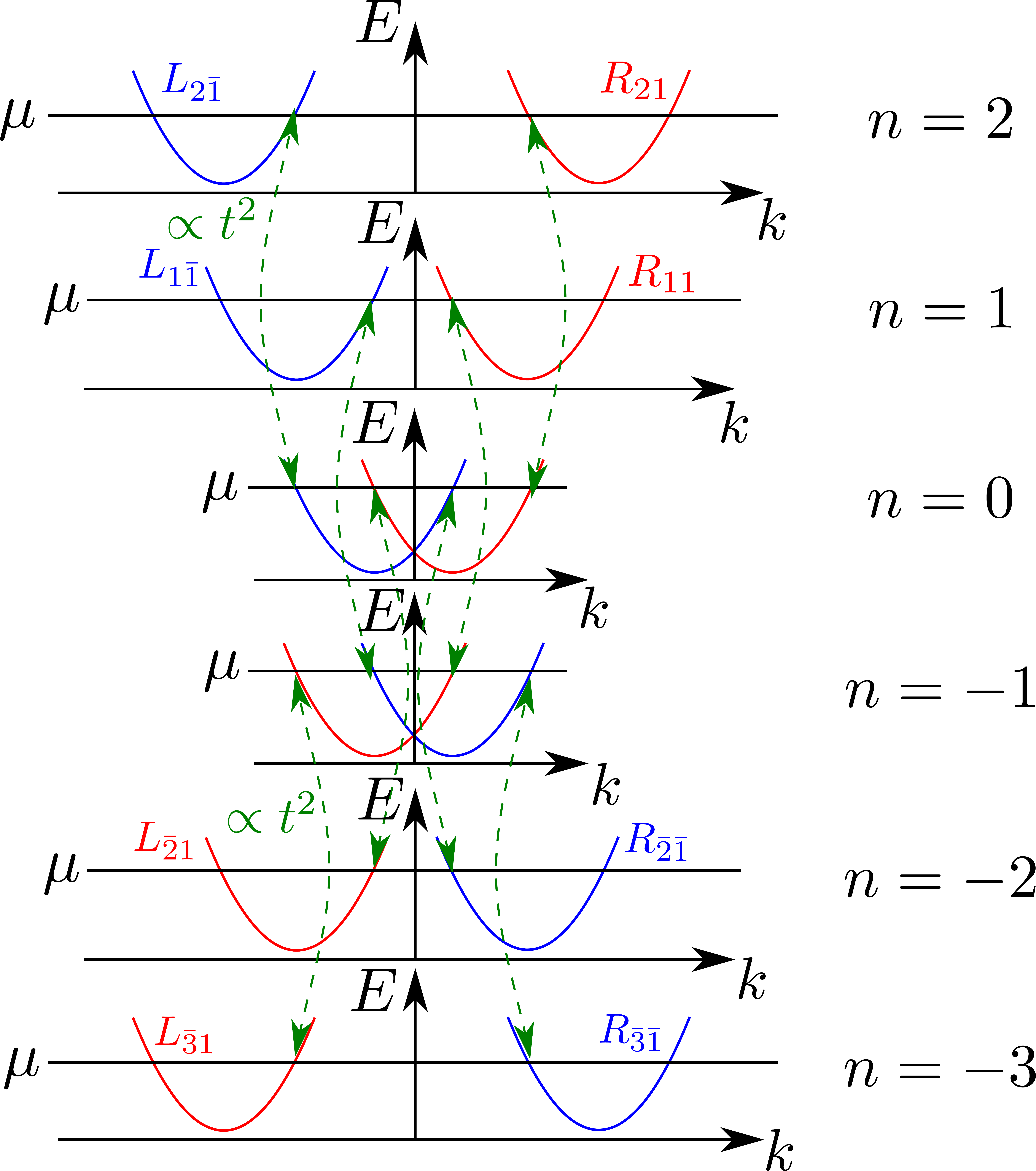}
\caption{The spectrum of a strip consisting of 6 stripes in the spin Hall regime characterized by the filling factor $\nu=2$. The chemical potential $\mu$ is uniform in the stripe and is kept to be the same as in the the $\nu=1$ regime, $\mu=\mu_1$, see Fig. \ref{ISHE1}. The gradient of the Rashba SOI leads to a shift of Fermi wavevectors $k_{Fn\pm}$ from stripe to stripe such that $k_{Fn+}=k_{F(n+2)-}$. As a result, the tunneling $t$ between stripes is resonant in second-order of perturbation theory, and couples right $R_{n \sigma}$ and left $L_{(n+2\sigma)\sigma}$ movers. Consequently, the system is gapped in the bulk. However, there are four uncoupled modes left at the upper ($R_{21}$, $L_{2\bar1}$, $R_{11}$, and $L_{1\bar1}$) and lower ($L_{\bar3 1}$, $R_{\bar 3\bar1}$, $L_{\bar3 1}$, and $R_{\bar 3\bar1}$) edges. These two helical pairs of edge modes carry opposite spins in opposite directions, which corresponds to  the QSHE at $\nu=2$.}
\label{ISHE2_new}
\end{figure}

{\it Edge modes in $x$ direction.} Next, we focus on the edge modes propagating in the $x$ direction. These modes are localized at $y=0$ (lower edge of the strip, the $n_l$th stripe) and at $y=L_y$ (upper edge of the strip, the $n_u$th stripe). We again represent electron operators $\Psi_{n\sigma}$ in terms of slowly-varying right-mover $R_{n\sigma}(x)$  and left-mover $L_{n\sigma}(x)$  fields as
\begin{align}
&\Psi_{n1}(x) = R_{n1}(x) e^{i k_{Fn+}x} + L_{n1}(x) e^{i k_{Fn-}x},\nonumber\\
&\Psi_{n\bar 1}(x) = L_{n\bar1}(x) e^{-i k_{Fn+}x} + R_{n\bar1}(x) e^{-i k_{Fn-}x}.
\end{align}
The kinetic part of the Hamiltonian becomes
\begin{align}
&H_{0}= \sum_{n=n_l}^{n_u} \sum_{\sigma=\pm1} -i\hbar \upsilon_F\int dx\  \Big[R^\dagger_{n\sigma}(x) \partial_x   R_{n\sigma}(x) \\
&\hspace{140pt} -  L^\dagger_{n\sigma}(x) \partial_x   L_{n\sigma}(x) \Big] \nonumber.
\end{align}
The resonant hopping term between stripes in the regime $\nu=1$ is given by
\begin{align}
&H_t = t \sum_{n=n_l}^{n_u-1} \int dx\  \Big[L_{(n+1)1}^\dagger (x)  R_{n 1} (x)\nonumber\\
&\hspace{80pt}+ R_{(n+1)\bar 1}^\dagger (x)  L_{n\bar 1} (x) +H.c.\Big].
\end{align}
This term accounts for the coupling between left and right movers at two neighbouring stripes, see Fig. \ref{ISHE1}.  Importantly, the right (left) mover with spin up (down) of the $n$th stripe is coupled to the the left (right) mover with spin up (down) of the $(n+1)$th stripe. As a consequence, the bulk spectrum is fully gapped.

Next, we show that despite the gap there are localized edge modes at the right and left edges of the strip.
As one can see, the two fields defined at $y=0$ (lower edge of the strip, $n_l$ stripe),  $R_{n_l\bar 1}(x)$ and $L_{n_l1}(x)$, do not enter  the tunneling term, and similarly for $R_{n_u1}(x)$ and $L_{n_u\bar 1}(x)$ defined at $y=L_y$ (upper edge of the strip, $n_u$ stripe). Thus, these fields correspond to the gapless modes propagating along the strip edges, see Fig. \ref{ISHE1}.
Importantly, there is a single pair of helical modes, modes in which opposite spins propagate in opposite directions, at each edge which shows that  the system is in the QSHE regime at the filling factor $\nu=1$.\cite{Kane_Mele_graphene,Hasan_review}

In a next step, we again generalize above result to other integer filling factors $\nu=m$.
For example, for $\nu=2$, the effective hopping term is written as
\begin{align}
&H_{t}^{(\nu=2)} = t^{(2)} \sum_{n=n_l}^{n_u-2} \int dx\  \Big[L_{(n+2)1}^\dagger (x)  R_{n1} (x)\nonumber\\
&\hspace{90pt}+ R_{(n+2)\bar 1}^\dagger (x)  L_{n\bar 1} (x) +H.c.\Big],
\end{align}
where the strength $t^{(2)} \propto t^2/E_F$ is determined in the second-order perturbation theory by considering the effective coupling as a result of two subsequent tunneling events.\cite{Stripes_arxiv} 
As follows directly from $H_{t}^{(\nu=2)}$ (see also Fig.~\ref{ISHE2_new}),  now there are two pairs of helical states at each of two edges. For example,  at $y=n_u$,  the following four modes: $R_{n_u1}(x)$, $R_{(n_u-1)1}(x)$, $L_{n_u\bar 1}(x)$, and $L_{(n_u-1)\bar 1}(x)$, stay gapless.
Two right propagating spin up modes and two left propagating spin down modes correspond to the spin Hall effect at the filling factor $\nu=2$. By analogy, this approach can be extended to other integer filling factors.

\subsection{Fractional Quantum Spin Hall Effect}

In this subsection, we focus on the QSHE in the fractional regime characterized by the filling factor $\nu=1/m$ with $m$ a positive odd integer. This regime can be achieved in a system where the gradient of SOI (and, consequently, the corresponding effective magnetic field) is $m$ times larger than in the case $\nu=1$ [see Eq. (\ref{grad})],
\begin{align}
\alpha_n^{(1/m)} = m(2n+1) \alpha_0.
\end{align} 
Here, we again keep the chemical potential to be constant at $\mu=\mu_1$.
In such systems the direct tunneling between right- and left-movers is not possible. As a result, the opening of gaps in the spectrum is possible only in the regime of strong electron-electron interactions when back-scattering terms begin to play a significant role.\cite{Kane_PRB,Kane_PRL,Stripes_PRL,Stripes_arxiv,oreg,Ady_FMF}
Below, without loss of generality, we focus on the regime of $\nu=1/3$ but the obtained results can be easily generalized  to other filling factors of the type $\nu=1/m$ with $m$ being an odd integer.

\begin{figure}[!tb]
\includegraphics[width=\linewidth]{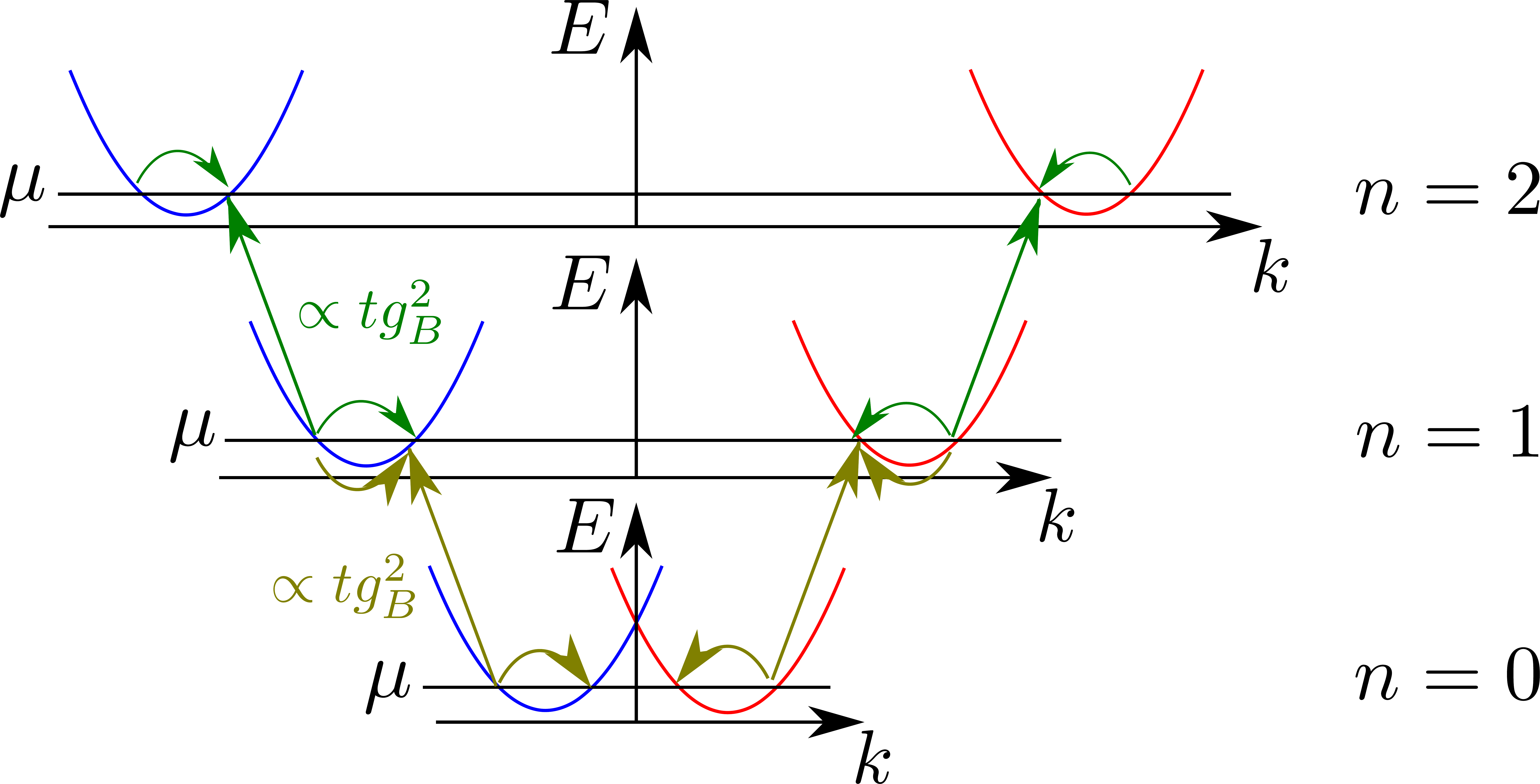}
\caption{The sketch of tunneling events in the system  brought in the QSHE regime at the filling factor $\nu=1/3$. The gradient of the Rashba SOI leads to a shift of Fermi wavevectors $k_{Fn\pm}$ from stripe to stripe such that $k_{F(n+1)+}=k_{Fn-}+4 k_{so}$. As a consequence, direct tunneling between stripes is forbidden by the momentum conservation law. However, if back-scattering terms induced by electron-electron interactions inside two neighbouring stripes are taken into account, the tunneling again becomes resonant and leads to the gapped bulk spectrum. Similarly to the integer case $\nu=m$, there are helical modes that propagate opposite spins in opposite directions. However, these modes transport fractional charges $e/3$.}
\label{ISHE1/3}
\end{figure}

First, we construct the tunneling term $H_{t}^{(\nu=1/3)}$ in leading order that conserves both momentum and spin (see also Fig. \ref{ISHE1/3}),
\begin{widetext}
\begin{align}
&H_{t}^{(\nu=1/3)} = \frac{g_t}{2} \sum_{n=n_l}^{n_u-1}  \int dx\  \Big([L_{(n+1)1}^\dagger (x)  R_{n1} (x)][L_{(n+1)1}^\dagger (x)  R_{(n+1)1} (x)][L_{n1}^\dagger (x)  R_{n1} (x)]\nonumber\\
&\hspace{140pt}+[R_{(n+1)\bar 1}^\dagger (x)  L_{n\bar 1} (x)][R_{(n+1)\bar 1}^\dagger (x)  L_{(n+1)\bar 1} (x)][R_{n\bar 1}^\dagger (x)  L_{n\bar 1} (x)]   +H.c.\Big).
\end{align}
\end{widetext}
Here, $g_t$ is proportional to the initial tunneling matrix element $t$ and to $g_B^2$, where $g_B$ describes the strength of the back-scattering term arising from electron-electron interactions.  As pointed out in previous work,\cite{Kane_PRB,Kane_PRL,oreg} it is too challenging to take into account all possible scattering terms and to solve the corresponding RG equations for Luttinger liquid parameters.  From now on, we just assume that $H_{t}^{(\nu=1/3)}$ is the most relevant term among all terms satisfying the spin and momentum conservation laws. This means that either the scaling dimension $K_t^{(1/3)}$ of $H_{t}^{(\nu=1/3)}$ is both smaller than two,  $K_t^{(1/3)}<2$, such that $g_t$ grows at low energy and, thus, the term is relevant, and smaller than the scaling dimensions of competing terms; or the bare $g_t$ is of order unity, such that $g_t$ does not flow as the perturbative RG treatment is not applicable.

Second,  to treat electron-electron interactions in one-dimensional systems, we switch to the  Luttinger liquid formalism based on bosonization. For this, we introduce chiral fields 
$\phi_{rn\sigma}$, 
defined via
\begin{align}
R_{n\sigma}(x) = e^{i \phi_{1n\sigma}(x)}\,  , \;\;\; L_{n\sigma}(x) = e^{i \phi_{\bar 1n\sigma}(x)}.
\end{align}
The anticommutation relations between two different fermionic operators $R_{n\sigma}$  and $L_{n\sigma}$  are satisfied in the bosonic representation of  $\phi_{rn\sigma}$ via Klein factors, which we do not take into account explicitly in the present work. At the same time, the anticommutation relation for the same fermionic operator can be satisfied explicitly by choosing the following commutation relation for the corresponding bosonic fields as
\begin{align}
[\phi_{rn\sigma}(x),\phi_{rn\sigma}(x')]=ir\pi\ {\rm sgn}(x-x').
\label{comm_1}
\end{align}

The tunneling term $H_{t}^{(\nu=1/3)}$, rewritten in bosonized form, becomes
\begin{align}
&H_{t}^{(\nu=1/3)}=\sum_{n=n_l}^{n_u-1}  \\
& \Big[g_t \cos (2\phi_{\bar1(n+1)1}-2\phi_{1n1}+\phi_{\bar1n1}-\phi_{1(n+1)1})\nonumber\\
&\hspace{10pt}+ g_t \cos (2\phi_{1(n+1)\bar1}-2\phi_{\bar 1n\bar1}+\phi_{1n\bar1}-\phi_{\bar 1(n+1)\bar1})\Big].\nonumber
\end{align}
In a next step, we introduce new bosonic fields $\tilde\phi_{rn\sigma} =(2\phi_{rn\sigma}-\phi_{\bar rn\sigma})/3$ that obey non-trivial commutation relations\cite{Ady_FMF} 
\begin{align}
[\tilde\phi_{rn\sigma}(x),\tilde\phi_{rn\sigma}(x')]=(ir\pi/3)\ {\rm sgn}(x-x'),
\label{com_bos}
\end{align}
which follows directly from Eq. (\ref{comm_1}).
This leads to the simplified form of $H_{t}^{(\nu=1/3)}$,
\begin{align}
&H_{t}^{(\nu=1/3)}=\sum_{n=n_l}^{n_u-1}  \Big[g_t \cos [3(\tilde \phi_{\bar1(n+1)1}-\tilde \phi_{1n1})]\nonumber\\
&\hspace{80pt}+ g_t \cos [3(\tilde\phi_{1(n+1)\bar1}-\tilde\phi_{\bar 1n\bar1})]\Big].
\end{align}
Again, we note that $\tilde\phi_{1n_l\bar1}$ and $\tilde\phi_{\bar 1 n_l 1}$ defined at the lower strip edge as well as $\tilde\phi_{1n_u 1}$ and $\tilde\phi_{  \bar 1n_u\bar 1}$ defined at the upper strip edge do not enter  the tunneling term $H_{t}^{(\nu=1/3)}$, and therefore they stay gapless. Moreover, the elementary excitations at these edge modes are non-trivial as follows directly from the commutation relations between the fields \cite{Ady_FMF,oreg,Stripes_arxiv,Stripes_PRL,Kane_PRB,Kane_PRL} [see Eq. (\ref{com_bos})]. Hence, these counter-propagating edge modes with opposite spins carry the fractional charge $e/3$, and the system is, thus, in  the QSHE regime at  filling factor $\nu=1/3$.

\vspace{20pt}

\section{Quantum Spin Hall Effect in Bilayer systems}

\subsection{Model}

\begin{figure}[!bT]
\includegraphics[width=\linewidth]{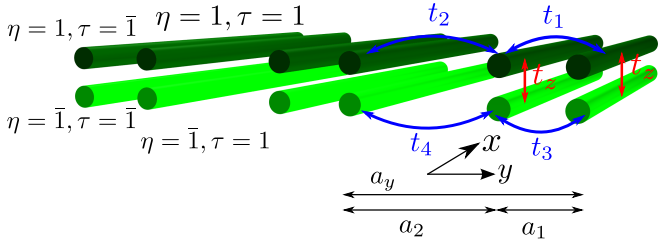}
\caption{The system consist of two strip of stripes aligned in the $y$ direction. The unit cell of size $a_y$ consists of four stripes: two belong to the upper strip and two to the lower strip. Each stripe in the system is characterized by three indices $(n\eta\tau)$, where $n$ denotes the position of the unit cell, $\eta=\pm1$ denotes the position in the upper or lower strip, and $\tau=\pm 1$ denotes the left/right stripe in the given unit cell. The two strips have effective masses with opposite signs. In addition, there is a uniform SOI in the $x$ direction inside each stripe. The upper and lower strips are coupled via spin-preserving tunneling $t_z$, whereas spin is flipped during the tunneling process $t_{i}$ between stripes inside the same strip that corresponds to the SOI effects affecting the motion in the $y$ direction. In particular, we are interested in the configuration where the distances between stripes inside the unit cell $a_{1}$ ($a_{2}$) corresponds to the spin-orbit length in the $y$ 
direction in the upper (lower) strip, and $a_{1}\neq a_{2}$, such that $t_{1}>t_{2}$ and $t_{4}>t_{3}$.  }
\label{model_bilayer}
\end{figure}

In this section we focus on bilayer systems that allows us to work with uniform systems and to avoid gradients in the system parameters. To be more specific, we consider a setup composed of an upper strip ($\eta=1$) with a positive mass $m$ and of a lower strip ($\eta=\bar 1$) with a negative mass $-m$. The unit cell  is of size $a_y$ and consists of two stripes in the upper/lower strip, labeled by the index $\tau=\pm 1$ separated by distances $a_{1}$ and $a_{2}$, $a_y=a_{1}+a_{2}$.

\begin{figure}[!tb]
\includegraphics[width=\linewidth]{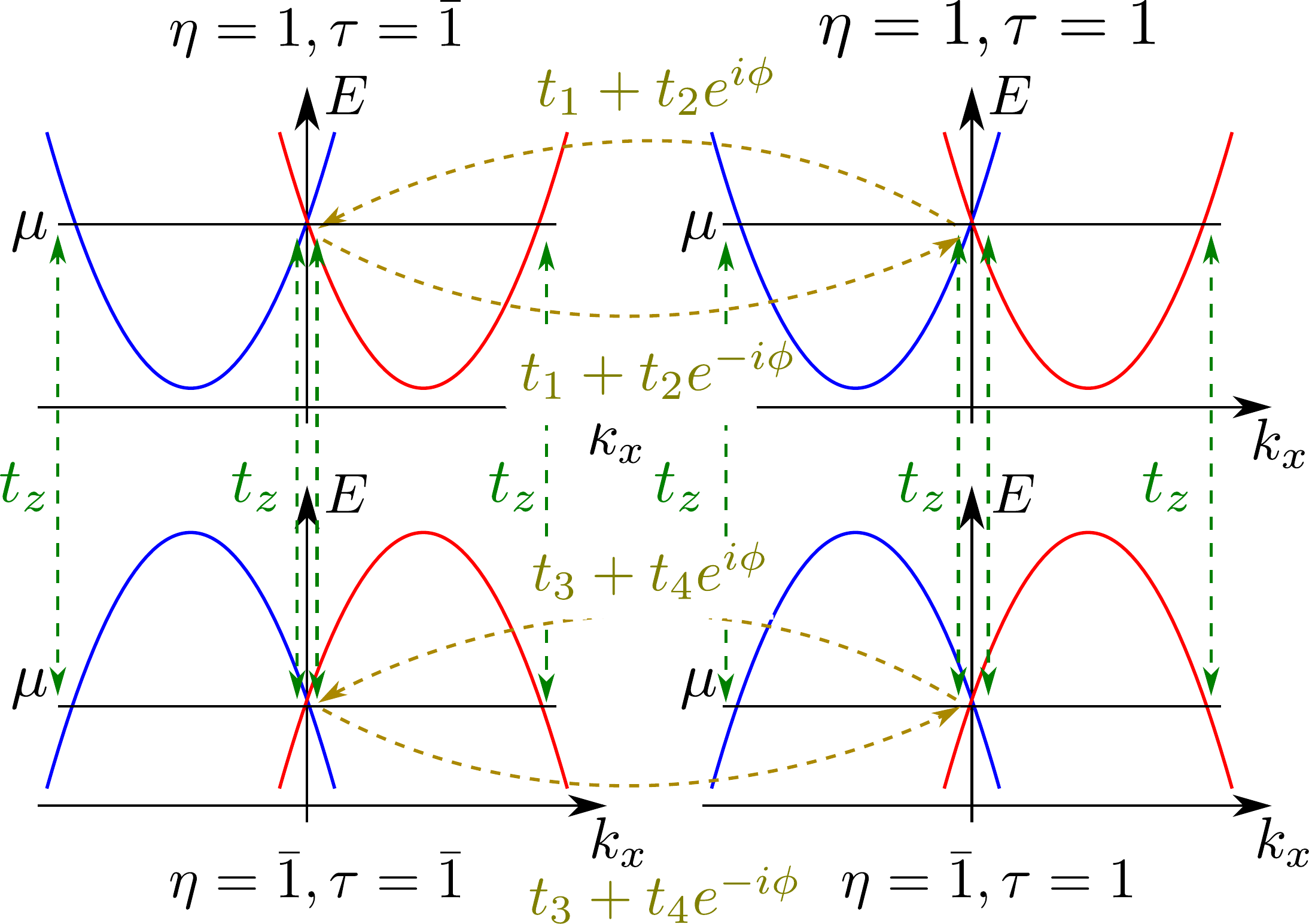}
\caption{The energy spectrum of the bilayer system (see Fig.~\ref{model_bilayer}). The upper (lower) stripes have a positive (negative) mass $m$ and a positive (negative) SOI in the $x$ direction. The spin-conserving interstrip tunneling terms of strength $t_z$ determine the size of the gap at  finite momenta (gaps are not shown), whereas at zero momentum they compete in the opening of gaps  with the spin-flipping intrastrip tunneling amplitudes of strength $t_{i}$ with $i=1,2,3,4$. If the intrastrip tunneling dominates, the system hosts edge states.
}
\label{spectrum_bilayer}
\end{figure}

The kinetic term in the Hamiltonian assumes the form
\begin{align}
&H_{0}= \sum_{\sigma,\tau,\eta=\pm1} \sum_n\int dx\ \nonumber\\
&\hspace{35pt}\Psi_{n\eta\tau\sigma}^\dagger (x)  \left(-\eta\frac{\hbar^2\partial_x^2}{2m} -\eta \mu \right)\Psi_{n\eta\tau\sigma} (x),
\end{align}
where the annihilation operator $\Psi_{n\eta\tau\sigma}(x)$ removes an electron with the spin $\sigma=\pm1$ at the point $x$ of the $(\eta\tau)$-stripe in the $n$th unit cell. The electron density, as well as
chemical potential $\mu$, is uniform inside the two strips. The spin quantization axis $z$ is determined by the SOI, which is of strength $\alpha$ and acts along the stripes (in the $x$ direction). The corresponding SOI term in the Hamiltonian reads 
\begin{align}
&H_{SOI} = -i   \sum_{\sigma,\sigma',\tau,\eta=\pm1} \sum_n\int dx\nonumber\\
&\hspace{70pt}\alpha \eta \Psi_{n\eta\tau\sigma}^\dagger (x) (\sigma_3)_{\sigma\sigma'} \partial_x \Psi_{n\eta\tau\sigma'} (x).
\end{align}
The dispersion relation of the ($\eta\tau$)-stripe (see Fig. \ref{spectrum_bilayer}) close to the common chemical potential is given by
\begin{align}
E_{\eta\tau\sigma} = \eta\left[\frac{\hbar^2(k-\sigma k_{so})^2}{2m}-\mu\right],
\end{align}
where we have absorbed a constant shift of energy in the chemical potential assuming the charge neutrality, and the SOI wavevector $k_{so}$ is given by $k_{so}= m \alpha/\hbar^2$.

The interstrip tunneling is spin-conserving,
\begin{align}
&H_z =t_z \sum_{\sigma,\tau=\pm1} \sum_n \int dx\ \Big[\Psi_{n 1 \tau \sigma}^\dagger\Psi_{n \bar1 \tau \sigma}+H.c.\Big].
% \nonumber\\
% &\hspace{30pt}+
% \Psi^\dagger_{n \bar1 \tau \sigma}\Psi_{n 1 \tau \sigma}\Big].
\end{align}
In contrast to that, the spin gets flipped for intrastrip tunnelings,
\begin{align}
&H_{y1\eta}= t_{y1\eta}  \sum_n \int dx\ \Big[\Psi_{n \eta 1 1}^\dagger\Psi_{n \eta \bar 1 \bar 1}\\
&\hspace{135pt}-\Psi_{n \eta 1  \bar 1}^\dagger\Psi_{n \eta \bar 1  1}+ H.c.\Big],\nonumber\\
&H_{y2\eta}= t_{y2\eta}   \sum_n \int dx\ \Big[\Psi_{(n+1) \eta \bar 1 1}^\dagger\Psi_{n \eta  1 \bar 1}\\
&\hspace{100pt}-\Psi_{(n+1) \eta \bar 1 \bar 1}^\dagger \Psi_{n \eta  1  1}+H.c.\Big],\nonumber
\end{align}
where $t_{y11}=t_1$, $t_{y1\bar 1}=t_3$, $t_{y21}=t_2$, and $t_{y2\bar 1}=t_4$. We could have also included spin-conserving tunneling acting in the $y$ direction, however, it couples right (left) movers with right (left) movers and, thus, does not lead to the gaps in the spectrum but just distorts the dispersion relation in the $x$ direction.
Hence, to keep the system simple, we neglect such tunneling terms.

\subsection{ Integer Quantum Spin Hall Effect}

We again begin with the non-interacting model and demonstrate that the system can be brought into the spin Hall regime that is characterized by the presence of two edge modes with opposite spins propagating in opposite directions, which can be identified with the QSHE regime at filling factor $\nu=1$.\cite{Kane_Mele_graphene,Hasan_review} The chemical potential $\mu$ is tuned to the SOI energy, {\it i.e.} at the crossing point between spin up and spin down branches at $k=0$, see Fig. \ref{spectrum_bilayer}.

We also switch from the fermionic operators $\Psi_{n \eta \tau \sigma}(x)$ to the slowly-varying right- [$R_{n \eta \tau \sigma}(x)$] and left- [$L_{n \eta \tau \sigma}(x)$]  mover fields,
\begin{align}
\Psi_{n \eta \tau \sigma}(x) = e^{ik_{F 1 \eta \tau \sigma} x} R_{n \eta \tau \sigma}(x) + e^{-ik_{F \bar 1 \eta \tau \sigma} x} L_{n \eta \tau \sigma}(x),
\end{align}
where the corresponding Fermi vectors $k_{F r \eta \tau \sigma}$ are given by $k_{F\sigma 1 \tau\sigma}=2\sigma k_{so}$, $k_{F\bar \sigma \bar1 \tau \sigma}=2\sigma k_{so}$, and $k_{F\bar \sigma 1\tau\sigma}=k_{F \sigma \bar 1\tau\sigma}=0$.

{\it Edge modes in $y$ direction.}
Similarly, to the previous section, we Fourier transform the total Hamiltonian  $H=H_0+H_{SOI}+H_z+\sum_{\eta=\pm1} (H_{y1\eta}+H_{y2\eta})$ to $k_y$-momentum space and work with the basis composed of $R_{k_y\eta\tau\sigma}$ and  $L_{k_y\eta\tau\sigma}$: $\Psi_{k_y}$= $(R_{k_y111}$, $L_{k_y111}$, $R_{k_y11\bar1}$, $L_{k_y11\bar1}$, $R_{k_y1\bar11}$, $L_{k_y1\bar11}$, $R_{k_y1\bar1\bar1}$, $L_{k_y1\bar1\bar1}$, $R_{k_y\bar111}$, $L_{k_y\bar111}$, $R_{k_y\bar11\bar1}$, $L_{k_y\bar11\bar1}$, $R_{k_y\bar1\bar11}$, $L_{k_y\bar1\bar11}$, $R_{k_y\bar1\bar1\bar1}$, $L_{k_y\bar1\bar1\bar1})$. The Hamiltonian density $\mathcal H$, $H=\int dx\ \Psi^\dagger_{k_y} \mathcal H \Psi_{k_y}$, is written in terms of Pauli matrices as
\begin{align}
 &\mathcal H=\hbar \upsilon_F \hat k \lambda_3 + t_z \tau_1 \lambda_1 \\
 &\hspace{20pt}- [t_{1}-t_{2} \cos (k_ya_y)] (1+\tau_3)\eta_2 (\sigma_2\lambda_1 - \sigma_1 \lambda_2)/4\nonumber\\
 &- t_{2} \sin (k_ya_y) (1+\tau_3)\eta_1 (\sigma_2\lambda_1
 - \sigma_1 \lambda_2)/4 \nonumber\\
 &\hspace{20pt}- [t_{3}-t_{4} \cos (k_ya_y)] (1-\tau_3)\eta_2 (\sigma_2\lambda_1 + \sigma_1 \lambda_2)/4 \nonumber\\
 &+t_{4} \sin (k_ya_y)(1-\tau_3)\eta_1 (\sigma_2\lambda_1 + \sigma_1 \lambda_2)/4 \nonumber.
\end{align}
The Pauli matrices $\lambda_i$ ($\sigma_i$) act on right/left-mover (spin up/down) space, whereas the Pauli matrices $\tau_i$ ($\eta_i$) act on the first/second stripe in the unit cell (upper/lower strip) space.
We note that the system belongs to the DIII topological class with the time-reversal operator $U_T$ given by $U_T=\sigma_2 \lambda_1$ and with the chiral symmetry operator $U_C$ given by $U_C = \lambda_3$.\cite{Neupert}

\begin{figure}[!tb]
\includegraphics[width=\linewidth]{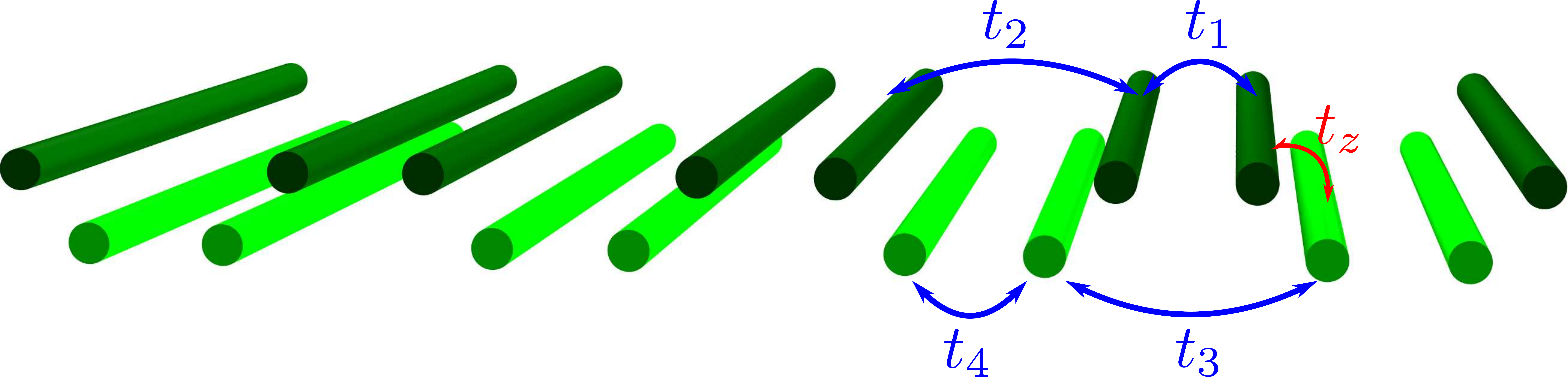}
\caption{A setup formed by two coupled strips which consists of stripes stacked in an armchair-type of order (compare with Fig. \ref{model_bilayer}). The tunneling amplitudes $t_1$ and $t_4$ that correspond to  short distance tunneling between stripes are naturally larger than the tunneling amplitudes $t_2$ and $t_3$ that correspond to  large distance tunneling between stripes.}
\label{armchair}
\end{figure}

For the sake of simplicity, we focus on the case when  $t_{1}=t_{4}>t_{2}=t_{3} \geq 0$. This can be achieved at least in two different configurations.
In the first setup, the distance between stripes in the unit cell $a_{i}$ equals to the spin orbit length in the $y$ direction of one of the two strips, see Fig, \ref{model_bilayer}. For example, if $a_{1}$ ($a_{2}$) is the SOI length in the upper (lower) strip, when the spin-flipping tunneling amplitude $t_{1}$ ($t_{4}$) dominates over $t_{2}$ ($t_{3}$) provided that $a_{1}$ and $a_{2}$ are substantially different and not commensurable. In the second setup, the stripes are stacked in a so-called armchair-type order (see Fig. \ref{armchair}), in analogy to graphene edges. In this case, $t_1$ and $t_4$ are the largest tunneling amplitudes as they correspond to the tunneling between pairs of the  closest stripes.

\begin{figure}[!b]
\includegraphics[width=0.8\linewidth]{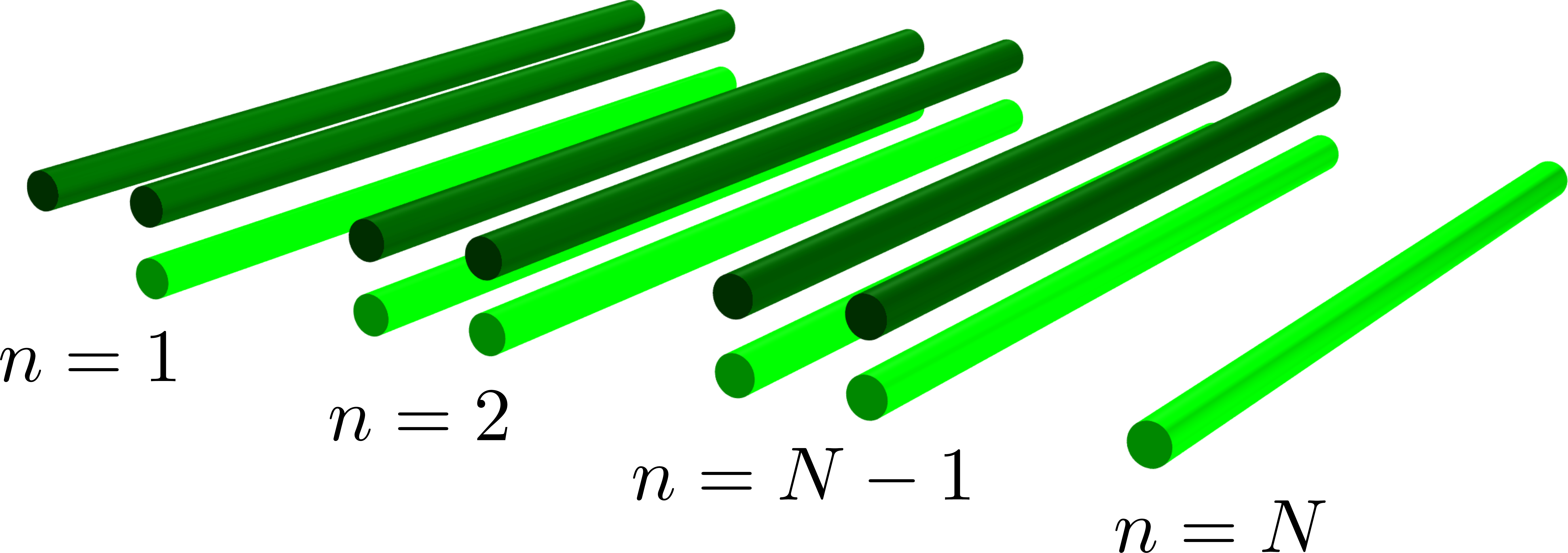}
\caption{The system of two coupled strips finite in the $y$ direction (see also Figs.~\ref{model_bilayer}). The $(\bar1\bar1)$-stripe is missing in the first ($n=1$) unit cell, whereas in the last ($n=N$) unit cell it is the only present stripe.}
\label{boundary}
\end{figure}

The energy spectrum is given by
\begin{align}
&E_{1\pm}^2=(\hbar \upsilon_F k)^2+t_{z}^2,\\
&E_{2\pm\pm}^2=\big[(\hbar \upsilon_F k)^2+t_{z}^2 + t_{3}^2+t_{4}^2 - 2t_{3}t_{4}\cos (k_ya_y) \nonumber\\
&\hspace{60pt} \pm 2 t_z (t_{3}+t_{4}) \sin{(k_ya_y)\big]/2},
\end{align}
where the level $E_{1\pm}$  ($E_{2\pm\pm}$) is fourfold (twofold) degenerate. 
The system is gapless at $k_ya_y=\pm \pi$ if $ t_z =t_{3}+ t_{4}$, but, otherwise, gapped in the bulk for all $k_y$ momenta. This closing of the bulk gap signals the topological phase transition that separates the topological phase with edge modes from the trivial phase without edge modes. More concrete, if $t_z<t_{3}+ t_{4}$, there are edge modes propagating in the $y$ direction and localized in the $x$ direction.

As an illustrative example, we consider a particular case $t_{3}=0$. 
The edge mode spectrum is given by
\begin{align}
E_{\pm}=\pm t_z \cos (k_y a_y/2),
\end{align}
where $k_y a_y \in [0, 2\pi)$.
The corresponding wavefunctions are given in the basis composed of $\Psi_{\eta\tau\sigma}$, $(\Psi_{111}$, $\Psi_{11\bar1}$, $\Psi_{1\bar11}$, $\Psi_{1\bar 1\bar1}$, $\Psi_{\bar111}$ ,$\Psi_{\bar11\bar1}$, $\Psi_{\bar1\bar11}$, $\Psi_{\bar1\bar 1\bar1})$, by
\begin{align}
\Phi_+(x,y)=\begin{pmatrix}
e^{-x/\xi_1} e^{-ik_F x}-e^{-x/\xi_2}\\
0\\
0\\
i(e^{-x/\xi_1} e^{ik_F x} -e^{-x/\xi_2})\\
(e^{-x/\xi_1} e^{-ik_Fx}-e^{-x/\xi_2})e^{-ik_ya_y/2}\\
0\\
0\\
i (e^{-x/\xi_1} e^{ik_Fx}-e^{-x/\xi_2})e^{ik_ya_y/2}
\end{pmatrix} e^{-ik_y y}
\end{align}
for the $E_+$ branch.
On the other hand, for the $E_-$ branch,
they are given by 
\begin{align}
\Phi_-(x,y)=\begin{pmatrix}
0\\
e^{-x/\xi_1} e^{ik_F x}-e^{-x/\xi_2}\\
i(e^{-x/\xi_1} e^{-ik_F x} -e^{-x/\xi_2})\\
0\\
0\\
-(e^{-x/\xi_1} e^{ik_Fx}-e^{-x/\xi_2})e^{-ik_ya_y/2}\\
-i(e^{-x/\xi_1} e^{-ik_Fx}-e^{-x/\xi_2})e^{ik_ya_y/2}\\
0
\end{pmatrix} e^{-ik_y y}.
\end{align}
Here the localization lengths are given by $\xi_1=\hbar \upsilon_F/[t_z \sin (k_y a_y/2)]$ and $\xi_2=\hbar \upsilon_F/[t_{4} - t_z \sin (k_y a_y/2)]$.
The two edge modes $\Phi_+(x,y)$ and $\Phi_-(x,y)$ are Kramers partners and connected by the time-reversal operator given by $T=i\sigma_2  K$ with $K$ being the complex conjugation operator.

In contrast to the previous case where the edge modes were carrying a particular spin projection, the edge modes along the $y$ directions considered above carry an oscillating in space spin polarization that changes inside the unit cell. Given this property, it seems more appropriate to refer to such a system as to a topological insulator rather than as to a system in QSHE regime.\cite{Hasan_review}

{\it Edge modes in $x$ direction.} Next, we focus on the edge modes propagating along the $x$ direction. Again, for a sake of simplicity we consider a case when $t_{1}=t_{4}>t_{2}=t_{3}=0$. In addition, we  assume that the bilayer strip terminates at both ends with only one stripe instead of two, see Fig.~\ref{armchair} and Fig.~\ref{boundary}. For example, as shown in Fig.~\ref{boundary}, the first unit cell ($n=1$) misses ($\bar1 \bar1$)-stripe, and the last unit cell ($n=N$) contains only ($\bar1 \bar1$)-stripe. This allows us to simplify analytical calculations. Moreover, already knowing about the presence of the edge modes along the $y$-direction, we know that the presence of edge modes in the $x$ direction should not be sensitive to a particular choice of boundary conditions.

Rewriting tunneling terms $H_z$, $H_{y1\eta}$, and  $H_{y2\eta}$ in terms of right-mover $R_{n\eta\tau\sigma}(x)$ and left-mover $L_{n\eta\tau\sigma}(x)$ fields, we arrive at
\begin{align}
&H_z =t_z \sum_{n=1}^{N-1} \int dx\ \Big[R_{n 1 \tau \sigma}^\dagger L_{n \bar1 \tau \sigma}+L^\dagger_{n \bar1 \tau \sigma}R_{n 1 \tau \sigma}+H.c.\Big],\nonumber\\
&H_{y1}= t_{4} \sum_{n=1}^{N-1} \int dx\ \Big[L_{n 1 1 1}^\dagger R_{n 1 \bar 1 \bar 1}-R_{n 1 1  \bar 1}^\dagger L_{n 1 \bar 1  1} +H.c. \Big],\nonumber\\
&H_{y4}= t_{4} \sum_{n=1}^{N-1} \int dx\ \Big[R_{(n+1) \bar 1 \bar 1 1}^\dagger L_{n \bar 1  1 \bar 1}-L_{(n+1) \bar 1 \bar 1 \bar 1}^\dagger R_{n \bar 1  1  1}\nonumber\\
&\hspace{180pt}+H.c.\Big],
\end{align}
where we again drop all fast-oscillating contributions.
As one can note, there are two fields in each of two boundary unit cells ($n=1$ and $n=N$) that do not enter in the tunneling terms and thus correspond to gapless edge modes provided that the bulk is gapped out, $t_{4}>t_z$. In particular, the pair of helical modes comprised of  $R_{1 1 \bar 1  1}$ and $L_{1 1 \bar 1  \bar 1}$ ($R_{N \bar 1 \bar 1  \bar 1}$ and $L_{N \bar 1 \bar 1 1}$) in the first (last) unit cell. We note that again opposite spins propagate in opposite directions that corresponds to the QSHE definition.\cite{Kane_Mele_graphene,Hasan_review}

\subsection{Fractional Quantum Spin Hall Effect}

Finally, we extend the  model introduced above 
to fractional filling factors, in particular, to $\nu=1/3$.  Similarly to the previous model, the direct intrastrip tunneling is suppressed if the chemical potential is tuned down to $\mu_{1/3}=E_{so}/9$ with $E_{so} = \hbar^2k_{so}^2/2m$. The Fermi wavectrors $k_{F \sigma \eta \tau\sigma}$ are given by
$k_{F\sigma 1 \tau\sigma}=k_{F\bar \sigma \bar1 \tau \sigma}=4\sigma k_{so}/3$, $k_{F\bar \sigma 1\tau\sigma}=k_{F \sigma \bar 1\tau\sigma}=2\sigma k_{so}/3$.  In addition, we again assume either that $t_{2}=t_{3}=0$ or that the corresponding tunneling terms are irrelevant in the renormalization group approach. From now on, we focus on the case where the intrastrip tunneling terms are more relevant than the interstrip ones either due to their scaling dimensions or due to their bare strengths. 

In addition, we assume that the leading intrastrip tunneling terms that conserve both momentum and spin are
\begin{align}
&H_{y1}^{(1/3)}= \frac{g_{y1}}{2} \sum_{n=1}^{N-1} \int dx\ \\
&\Big[(L_{n  1 1 1}^\dagger R_{n  1 \bar 1  \bar 1})(L_{n 1 1  1}^\dagger R_{n  1  1  1})(L_{n  1 \bar 1  \bar 1}^\dagger R_{n  1 \bar 1  \bar 1})\nonumber\\
&\hspace{25pt}-(R_{n  1 1 \bar 1}^\dagger L_{n  1 \bar 1  1})(R_{n 1 1 \bar 1}^\dagger L_{n  1  1 \bar 1})(R_{n  1 \bar 1  1}^\dagger L_{n  1 \bar 1  1})+H.c.\Big],\nonumber\\
&H_{y4}^{(1/3)}= \frac{g_{y4}}{2}\sum_{n=1}^{N-1} \int dx\ \\
&\Big[(R_{(n+1) \bar 1 \bar 1 1}^\dagger L_{n \bar 1  1 \bar 1})(R_{(n+1) \bar 1 \bar 1 1}^\dagger L_{(n+1) \bar 1  \bar 1  1})(R_{n \bar 1  1 \bar 1}^\dagger L_{n \bar 1  1 \bar 1})\nonumber\\
&-(L_{(n+1) \bar 1 \bar 1 \bar 1}^\dagger R_{n \bar 1  1  1})(L_{(n+1) \bar 1 \bar 1 \bar 1}^\dagger R_{(n+1) \bar 1  \bar 1  \bar 1})(L_{n \bar 1  1  1}^\dagger R_{n \bar 1  1  1}) \nonumber\\
&\hspace{180pt}+ H.c.\Big], \nonumber
\end{align}
where $g_{y1} \propto t_{1} g_B^2$, $g_{y4} \propto t_{4} g_B^2$.  These two terms are the most relevant ones among all terms either due to their scaling dimensions, such that they grow under the RG flow fastest at low energy, or due to their bare strengths being of order of unity, such that they are not subjected to the RG flow. Similarly to previous work,\cite{Kane_PRB,Kane_PRL,oreg}  we leave a complete analysis of the RG flow of Luttinger liquid parameters for elsewhere, which are needed to estimate the scaling dimensions explicitly,  and work here under the assumption that the selected terms
are relevant in above sense.

The corresponding interstrip tunneling term that commutes with both $H_{y1}^{(1/3)}$ and $H_{y4}^{(1/3)}$, such that it can be ordered simultaneously with them, and thus can lead to gaps in the spectrum\cite{Ady_FMF} is given by
\begin{align}
&H_z^{(1/3)} =\frac{g_z}{2} \sum_{n=1}^{N-1} \int dx\ \\
&\Big[(L_{n 1 \tau \bar 1}^\dagger R_{n \bar1 \tau \bar 1})(L_{n 1 \tau \bar 1}^\dagger R_{n 1 \tau \bar 1})(L_{n \bar 1 \tau \bar 1}^\dagger R_{n \bar1 \tau \bar 1})   \nonumber\\
&\hspace{15pt}+(R_{n 1 \tau 1}^\dagger L_{n \bar1 \tau 1})(R_{n 1 \tau 1}^\dagger L_{n 1 \tau 1})(R_{n \bar1 \tau 1}^\dagger L_{n \bar1 \tau 1})+H.c.\Big], \nonumber
\end{align}
where $g_{z} \propto t_{z} g_B^2$.

To analyze the spectrum further, we switch to bosonic chiral fields $\phi_{n r \eta\tau\sigma}$ defined via
\begin{align}
R_{n\eta\tau\sigma} = e^{i \phi_{n 1 \eta\tau\sigma}}\ {\rm and}\ L_{n\eta\tau\sigma} = e^{i \phi_{n \bar 1 \eta\tau\sigma}}.
\end{align}
To satisfy the anticommutation relations for the same fermionic operator at different spatial points, we choose the following commutation relations for the corresponding bosonic fields
\begin{align}
&[\phi_{n r \eta\tau\sigma}(x),\phi_{n' r' \eta'\tau'\sigma'}(x')]\nonumber\\
&\hspace{40 pt}=ir\pi \delta_{nn'} \delta_{rr'} \delta_{\eta \eta'}\delta_{\tau \tau'}\delta_{\sigma \sigma'} {\rm sgn}\ (x-x').
\end{align}
The anticommutation relations between two different fermionic operators are satisfied by a proper choice of Klein factors,\cite{giamarchi_book} which we do not include explicitly in our calculations.

As a result, we arrive at the tunneling terms in the form
\begin{align}
&H_{y1}^{(1/3)}= g_{y1} \sum_{n=1}^{N-1} [\cos(2\phi_{n \bar 1 111}+\phi_{n \bar 1 1\bar 1\bar1}-2\phi_{n 1 1 \bar  1 \bar 1}-\phi_{n 1 1 1 1})\nonumber\\
&\hspace{20pt}-\cos(2\phi_{n  1 11 \bar 1}+\phi_{n  1 1\bar 1 1}-2\phi_{n \bar1 1 \bar  1 1}-\phi_{n \bar 1 1 1 \bar 1})],\\
&H_{y4}^{(1/3)}= g_{y4} \sum_{n=1}^{N-1} \\
&[\cos(2\phi_{(n+1)  1 \bar1\bar11}+\phi_{n  1 \bar 1 1\bar1}-2\phi_{n \bar 1 \bar 1   1 \bar 1}-\phi_{(n+1) \bar 1 \bar 1 \bar 1 1})\nonumber\\
&\hspace{20pt}-\cos(2\phi_{(n+1)  \bar 1 \bar1\bar1\bar1}+\phi_{n  \bar1 \bar 1 11}-2\phi_{n 1 \bar 1   1  1}-\phi_{(n+1)  1 \bar 1 \bar 1 \bar1})],\nonumber\\
&H_z^{(1/3)} =g_z\sum_{n=1}^{N-1} [\cos(2\phi_{n \bar 1 1\tau \bar 1}+\phi_{n \bar 1 \bar 1 \tau \bar1}-2\phi_{n 1 \bar 1 \tau \bar 1}-\phi_{n 1 1 \tau \bar 1})\nonumber\\
&\hspace{20pt}+\cos(2\phi_{n  1 1\tau  1}+\phi_{n 1 \bar 1 \tau 1}-2\phi_{n \bar 1 \bar 1 \tau  1}-\phi_{n \bar 1 1 \tau  1})].
\end{align}

We can simplify these expressions by introducing new fields,
\begin{align}
&\tilde \phi_{n r \eta\tau\sigma}=(2\phi_{n r \eta\tau\sigma} - \phi_{n \bar r \eta\tau\sigma})/3,\\
&[\tilde \phi_{n r \eta\tau\sigma}(x),\tilde \phi_{n' r' \eta'\tau'\sigma'}(x')]\nonumber\\
&\hspace{20 pt}=(ir\pi/3) \delta_{nn'} \delta_{rr'} \delta_{\eta \eta'}\delta_{\tau \tau'}\delta_{\sigma \sigma'} {\rm sgn}\ (x-x').\label{tilde_phi}
\end{align}
This allows us to rewrite the tunneling terms as 
\begin{align}
&H_{y1}= g_{y1} \sum_{n=1}^{N-1} \big(\cos[3(\tilde \phi_{n \bar 1 111}-\tilde \phi_{n 1 1 \bar  1 \bar 1})]\nonumber\\
&\hspace{100pt}-\cos[3(\tilde \phi_{n  1 1 1 \bar 1}-\tilde \phi_{n \bar1 1 \bar  1 1})]\big),\\
&H_{y4}= g_{y4} \sum_{n=1}^{N-1}  \big(\cos[3(\tilde \phi_{(n+1)  1 \bar1\bar11}-\tilde \phi_{n \bar 1 \bar 1   1 \bar 1})]\nonumber\\
&\hspace{80pt}-\cos[3(\tilde \phi_{(n+1)  \bar 1 \bar1\bar1\bar1}-\tilde \phi_{n 1 \bar 1   1  1})]\big),\\
&H_z =g_z \sum_{n=1}^{N-1}  \big( \cos[3(\tilde \phi_{n \bar 1 1\tau \bar 1}-\tilde \phi_{n 1 \bar 1 \tau \bar 1})]\nonumber\\
&\hspace{90pt}+\cos[3(\tilde \phi_{n  1 1\tau  1}-\tilde \phi_{n \bar 1 \bar 1 \tau  1})]\big).
\end{align}

Again, we see that a pair of fields defined in the first unit cell ($n=1$) and a pair of fields defined at the last ($n=N$) unit cell do not enter in the tunneling term in the Hamiltonian. More concrete, the right propagating mode with spin up $\tilde \phi_{ 1 1 1 \bar 1  1}$  and the left propagating mode with spin down $\tilde \phi_{1 \bar  1 1 \bar 1  \bar 1}$ belonging to the $(1\bar1)$-stripe of the first unit cell stay gapless. The same is true for the left propagating mode with spin up $\tilde \phi_{N \bar 1 \bar 1 \bar 1  1}$  and the right propagating mode with spin down $\tilde \phi_{N 1 \bar 1 \bar 1  \bar 1}$ belonging to the $(\bar 1\bar1)$-stripe of the $N$th unit cell. Hence, we deal with  opposite spins propagating in opposite directions that confirm our hypothesis of the system being in the QSHE regime. In addition, the edge modes carry the fractional charge $e/3$, and their excitations possess non-trivial Abelian braiding statistics determined from Eq.~(\ref{tilde_phi}).\cite{Ady_FMF,oreg,
Stripes_arxiv,Stripes_PRL,Kane_PRB,Kane_PRL}

\begin{figure*}[!th]
\includegraphics[width=0.5\linewidth]{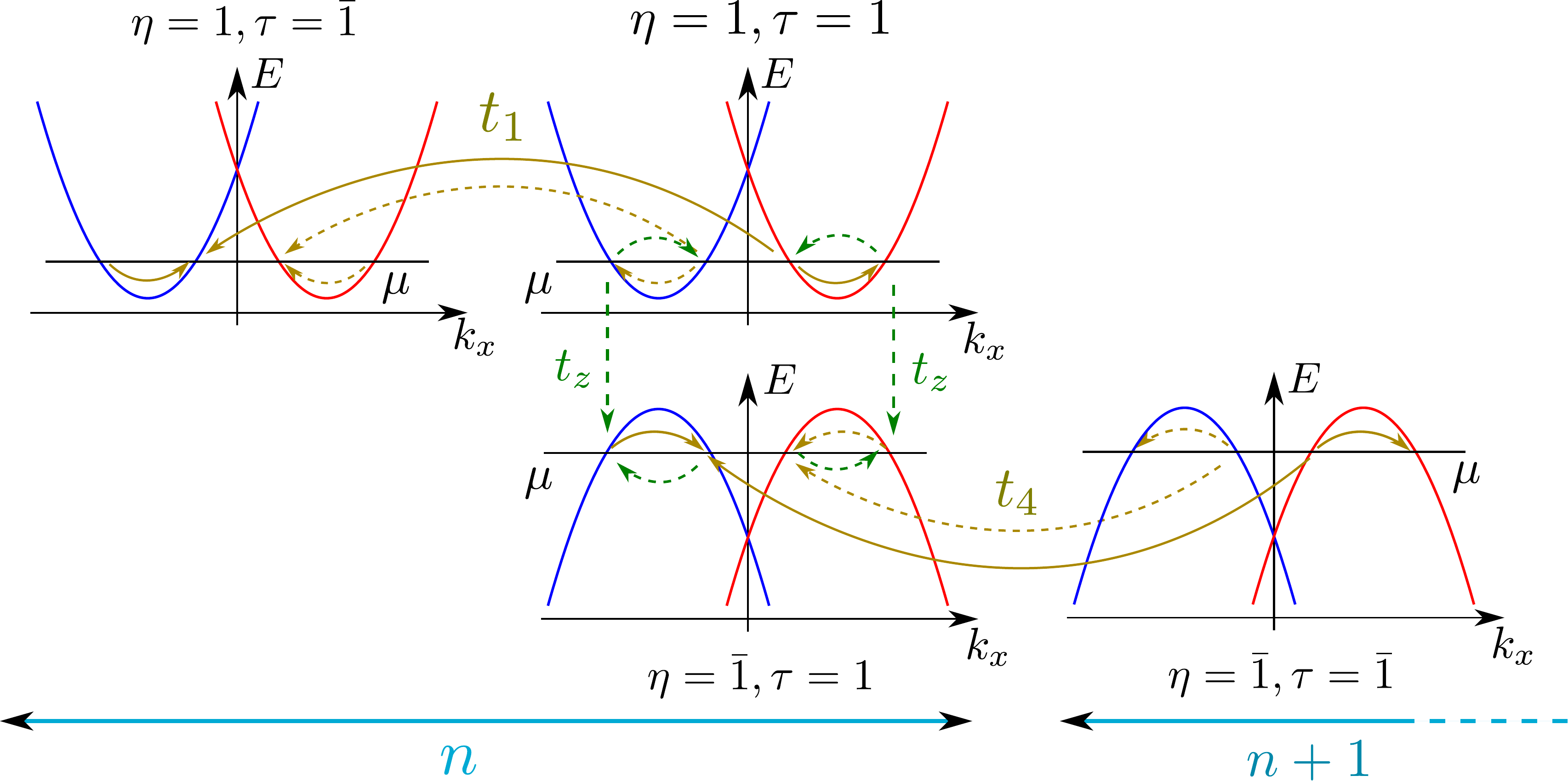}
\caption{The spectrum of the system of two coupled strips in the fractional regime at the filling factor $\nu=1/3$ (compare with Fig. \ref{spectrum_bilayer}). The chemical potential $\mu$ is tuned to $E_{so}/9$. The intrastrip tunneling (yellow arrows) is possible only if back-scattering terms are included. Consequently, interstrip tunneling terms (green arrows), which should commute with intrastrip tunneling terms, also involve back-scattering due to strong electron-electron interactions. }
\label{spectrum_FSHE}
\end{figure*}

\section{Conclusions}

We have constructed two strip of stripes models that exhibit the quantum spin Hall effect. The first one is based on a single layer consisting of a strip of stripes with a gradient of SOI pointing perpendicular to the stripes. The second one is a bilayer model composed of a tunnel-coupled pair of strips of stripes with reversed dispersion relation. This bilayer model can be effectively used for describing topological insulators originating from narrow band-gap semiconductors. \cite{TI_inverted,Patric_TI,Du_exp}
In the presence of strong electron-electron interactions, when back-scattering terms begin to play a crucial role, the system can be brought into the fractional quantum spin Hall regime. In this regime the charge carried by the modes is a fraction of the elementary electric charge, and excitations possess non-trivial Abelian braiding statistics.\cite{Hasan_review,Ady_FMF,oreg,Stripes_arxiv,Stripes_PRL,Kane_PRB,Kane_PRL}

In general, our work on strip of stripes models shows how by using ideas of representing two-dimensional systems as a system of  coupled one-dimensional channels\cite{Kane_PRB,Kane_PRL,Stripes_PRL,Stripes_arxiv,oreg,Neupert} we can generate new states of matter. The main advantage of such models is that we can use powerful theoretical methods such as Luttinger liquid description developed for one-dimensional system to address two-dimensional systems where such methods are absent. In addition, relying on the stability of topological states as long as the bulk gap is not closed, we can assume that the basic properties derived in the anisotropic limit in such strip of stripes models might stay valid also in the isotropic limit that for some cases is closer to experiments. Consequently, we can address several challenging questions raised in the field of  topological insulator such as, for example, the  role of nuclear spins or non-perfect quantization of conductance due to disorder.\cite{Konig,Konig_2,Amir_TI} In 
addition, it opens new ways to describe edge modes and braiding statistics of their exitations, which is of a great importance for Majorana fermion and parafermion physics.\cite{PF_Linder,Cheng,PF_TI_Amir}

\acknowledgments
We thank Daniel Loss and Mircea Trif for encouraging and helpful discussions. 
JK acknowledges funding from the Harvard Quantum Optics Center, and YT acknowledges funding from FAME (an SRC STARnet center sponsored by MARCO and DARPA).
This work was supported in part by the Kavli Institute for Theoretical Physics through Grant No. NSF PHY11-25915.

%%%%%%%%%%%%%%%%%%%%%%%


\begin{thebibliography}{99}




%Fractional Fermions



 \bibitem{Jackiw_Rebbi} R. Jackiw and C. Rebbi, Phys. Rev. D {\bf 13}, 3398 (1976).

 \bibitem{FracCharge_Su} W. P. Su, J. R. Schrieffer, and A. J. Heeger, Phys. Rev. Lett. {\bf 42}, 1698 (1979).
 
 \bibitem{FracCharge_Kivelson} S. Kivelson and J. R. Schrieffer, Phys. Rev. B {\bf 25}, 6447, (1982).


 
 \bibitem{CDW} S. Gangadharaiah, L. Trifunovic, and D. Loss, Phys. Rev. Lett. {\bf 108}, 136803 (2012).
 
 \bibitem{Two_field_Klinovaja_Stano_Loss_2012} J. Klinovaja, P. Stano, and D. Loss, Phys. Rev. Lett. {\bf 109}, 236801 (2012).
 
 \bibitem{Klinovaja_Loss_FF_1D} J. Klinovaja and D. Loss, Phys. Rev. Lett. {\bf 110}, 126402 (2013).
 
 \bibitem{FF_transport} D. Rainis, A. Saha, J. Klinovaja, L. Trifunovic, and D. Loss, Phys. Rev. Lett. {\bf 112}, 196803 (2014).

%Quantum Hall

\bibitem{Klitzing} K. v. Klitzing, G. Dorda, and M. Pepper, Phys. Rev. Lett. {\bf 45}, 494 (1980).

\bibitem{Tsui_82} D. C. Tsui, H. L. Stormer, and A. C. Gossard, Phys. Rev. Lett. {\bf 48}, 1559 (1982).

\bibitem{Halperin} B. I. Halperin, Phys. Rev. B {\bf 25}, 2185 (1982).

\bibitem{Laughlin_FQHI} R. B. Laughlin, Phys. Rev. Lett. {\bf 50}, 1395 (1983).

\bibitem{QHE_Review_Prange} R. E. Prange and S. M. Girvin, {\it The Quantum Hall Effect} (Springer, New York, 1990).

\bibitem{book_Jain} J. K. Jain, {\it Composite Fermions} (Cambridge University Press, Cambridge, 2007).


\bibitem{Lebed} A. G. Lebed, JETP Lett. {\bf 43}, 174 (1986).

\bibitem{Montambaux} D. Poilblanc, G. Montambaux, M. Heritier, and P. Lederer, Phys. Rev. Lett. {\bf 58}, 270 (1987).

\bibitem{Lebed_Gorkov} L. P. Gor'kov and A. G. Lebed, Phys. Rev. B {\bf 51}, 3285 (1995).

\bibitem{Yakovenko_PRB} V. M. Yakovenko, Phys. Rev. B {\bf 43}, 11353 (1991).

\bibitem{Lee_PRB}  D.-H. Lee, Phys. Rev. B {\bf 50}, 10788 (1994).

\bibitem{Yakovenko_review} V. M. Yakovenko, {\it The Physics of Organic Superconductors and Conductors} edited by A. G. Lebed, Springer series in Material Sciences {\bf 110},  529 (2008).

\bibitem{Kane_PRL} C. L. Kane, R. Mukhopadhyay, and T. C. Lubensky, Phys. Rev. Lett. {\bf 88}, 036401 (2001).

\bibitem{Kane_PRB} J. C. Y. Teo and C. L. Kane, Phys. Rev. B {\bf 89}, 085101 (2014).

\bibitem{Stripes_PRL} J. Klinovaja and D. Loss, Phys. Rev. Lett. {\bf 111}, 196401 (2013).

\bibitem{Stripes_arxiv} J. Klinovaja and D. Loss, arXiv:1305.1569.

\bibitem{Stripe_PRL_exp} K. Kobayashi, H. Satsukawa, J. Yamada, T. Terashima, and S. Uji, Phys. Rev. Lett. {\bf 112}, 116805 (2014).





%TI


\bibitem{Hasan_review} M. Z. Hasan and C. L. Kane, Rev. Mod. Phys. {\bf 82}, 3045 (2010).

\bibitem{Zhang_RMP} X. Qi and S. Zhang, Rev. Mod. Phys. {\bf 83}, 1057 (2011).

\bibitem{Volkov_TI1} B. A. Volkov and O. A. Pankratov, JETP Lett. {\bf 42}, 178 (1985).

\bibitem{Volkov_TI2} O. A. Pankratov, S. V. Pakhomov, and B. A. Volkov, Solid State Commun. {\bf 61}, 93 (1987).

\bibitem{Volkov_TI3} O. A. Pankratov and B. A. Volkov, in Landau Level Spectroscopy, edited by E. I. Rashba and G. Landwehr (North-Holland, Amsterdam, 1991), Chap. 14, p. 817.

\bibitem{Kane_Mele_graphene} C. L. Kane and E. J. Mele, Phys. Rev. Lett. {\bf 95}, 226801 (2005).

\bibitem{Bernevig} B. A. Bernevig, T. L. Hughes, and S.-C. Zhang, Science {\bf 314}, 1757 (2006).

\bibitem{Fu_Kane} L. Fu, C. L. Kane, and E. J. Mele, Phys. Rev. Lett. {\bf 98}, 106803 (2007).

\bibitem{TI_inverted} C. Liu, T.L. Hughes, X.-L. Qi, K. Wang, and S.-C. Zhang, Phys. Rev. Lett. {\bf 100}, 236601 (2008).

\bibitem{Zhang_TI} X.-L. Qi, T. L. Hughes, and S.-C. Zhang, Phys. Rev. B {\bf 78}, 195424 (2008).

\bibitem{TI_Ady} M. Levin and A. Stern, Phys. Rev. Lett. {\bf 103}, 196803 (2009).

\bibitem{Chamon_Mudry_2011} T. Neupert, L. Santos, C. Chamon, and C. Mudry, Phys. Rev. Lett. {\bf 106}, 236804 (2011).

\bibitem{Patric_TI} P. Michetti, J. C. Budich, E. G. Novik, and P. Recher, Phys. Rev. B 85, 125309 (2012).

\bibitem{Carlos_TI}	S. I. Erlingsson and J. C. Egues, arXiv:1312.2034.

\bibitem{Ady_wires} I. Seroussi, E. Berg, and Y. Oreg,  arXiv:1401.2671.

\bibitem{oreg} E. Sagi and Y. Oreg, arXiv:1403.1791.

\bibitem{Neupert} T. Neupert, C. Chamon, C. Mudry, and R. Thomale, arXiv:1403.0953.

\bibitem{Konig_2} M. Konig, S. Wiedmann, C. Brune, A. Roth, H. Buhmann, L. W. Molenkamp, X. Qi, and S. Zhang, Science
{\bf 318}, 766 (2007).

\bibitem{Konig}  M. Konig, H. Buhmann, L. W. Molenkamp, T. Hughes, C. Liu, X. Qi, and S. Zhang, J.  Phys. Soc. Jpn. {\bf 77}, 031007 (2008).

\bibitem{Roth_TI} A. Roth, C. Brüne, H. Buhmann, L. W. Molenkamp, J. Maciejko, X. Qi, and S. Zhang, Science {\bf 325}, 294 (2009).

\bibitem{Du_exp} I. Knez, R. R. Du, and G. Sullivan, Phys. Rev. Lett. {\bf 107}, 136603 (2011).

\bibitem{exp_1} J. Wang, H. Li, C. Chang, K. He, J. Lee, H. Lu, Y. Sun, X. Ma, N. Samarth, S. Shen, Q. Xue, M. Xie, and M. Chan, Nano Res. {\bf 5}, 739 (2012).

\bibitem{Nowack_TI} K. C. Nowack, E. M. Spanton, M. Baenninger, M. Konig, J. R. Kirtley, B. Kalisky, C. Ames, P. Leubner, C. Brune, H. Buhmann, L. W. Molenkamp, D. Goldhaber-Gordon, and K. A. Moler,
% mIaging currents in hgte quantum wells in the quantum spin hall regime.
 arXiv:1212.2203.
 
\bibitem{Amir_TI} S. Hart, H. Ren, T. Wagner, P. Leubner, M. Muhlbauer, C. Brune, H. Buhmann, L. W. Molenkamp, and A. Yacoby, arXiv:1312.2559.





%Majorana fermions citations

\bibitem{Read_2000} N. Read and D. Green, Phys. Rev. B {\bf 61}, 10267 (2000).

\bibitem{fu} L. Fu and C. L. Kane, Phys. Rev. Lett. {\bf 100}, 096407 (2008).

\bibitem{Nagaosa_2009} Y. Tanaka, T. Yokoyama, and N. Nagaosa, Phys. Rev. Lett. {\bf 103}, 107002 (2009).

\bibitem{Sato}M. Sato and S. Fujimoto, Phys. Rev. B {\bf 79}, 094504 (2009).


\bibitem{lutchyn_majorana_wire_2010} R. M. Lutchyn, J. D. Sau, and S. Das Sarma,
Phys. Rev. Lett. {\bf 105}, 077001 (2010).

\bibitem{oreg_majorana_wire_2010} Y. Oreg, G. Refael, and F. von Oppen,
Phys. Rev. Lett. {\bf 105}, 177002 (2010).

\bibitem{alicea_majoranas_2010} J. Alicea, Phys. Rev. B {\bf 81}, 125318 (2010).

\bibitem{demler_2011} L. Jiang, T. Kitagawa, J. Alicea, A. Akhmerov, D. Pekker, G. Refael, J. I. Cirac, E. Demler, M. D. Lukin, and P. Zoller, Phys. Rev. Lett. {\bf 106}, 220402 (2011).
%Majorana Fermions in Equilibrium and in Driven Cold-Atom Quantum Wires

\bibitem{MF_ee_Suhas} S. Gangadharaiah, B. Braunecker, P. Simon, and D. Loss, Phys. Rev. Lett. {\bf 107}, 036801 (2011).

\bibitem{potter_majoranas_2011} A. C. Potter and P. A. Lee, Phys. Rev. B {\bf 83}, 094525 (2011).

\bibitem{Klinovaja_CNT} J. Klinovaja, S. Gangadharaiah, and D. Loss, Phys. Rev. Lett. {\bf 108}, 196804 (2012).

\bibitem{bilayer_MF_2012} J. Klinovaja, G. J. Ferreira, and D. Loss, Phys. Rev. B  {\bf 86}, 235416 (2012).



\bibitem{RKKY_Basel} J. Klinovaja, P. Stano, A. Yazdani, and D. Loss, Phys. Rev. Lett. {\bf 111}, 186805 (2013).

\bibitem{RKKY_Simon} B. Braunecker and P. Simon, Phys. Rev. Lett. {\bf 111}, 147202 (2013).

\bibitem{RKKY_Franz} M. Vazifeh and M. Franz,  Phys. Rev. Lett. {\bf 111}, 206802 (2013).

\bibitem{MF_nanoribbon} J. Klinovaja and D. Loss, Phys. Rev. X {\bf 3}, 011008 (2013).

\bibitem{MF_Bena} C. Dutreix, M. Guigou, D. Chevallier, and C. Bena, arXiv:1309.1143.

\bibitem{MF_MOS} J. Klinovaja and D. Loss, Phys. Rev. B {\bf 88}, 075404 (2013).

%=-------------

\bibitem{Ando}
S. Sasaki, M. Kriener, K. Segawa, K. Yada, Y. Tanaka, M. Sato, and Y. Ando, 
Phys. Rev. Lett. {\bf 107}, 217001 (2011). 

\bibitem{mourik_signatures_2012}
V. Mourik, K. Zuo, S. M. Frolov, S. R. Plissard, E. P. A. M. Bakkers, and L. P. Kouwenhoven, Science, {\bf 336}, 1003 (2012). 

\bibitem{deng_observation_2012} 
M. T. Deng, C. L. Yu, G. Y. Huang, M. Larsson, P. Caroff, and H. Q. Xu, Nano Lett. {\bf 12}, 6414 (2012).

\bibitem{das_evidence_2012}
A. Das, Y. Ronen, Y. Most, Y. Oreg, M. Heiblum, and H. Shtrikman, Nat. Phys. {\bf 8}, 887 (2012). 

\bibitem{Rokhinson} L. P. Rokhinson, X. Liu, and J. K. Furdyna, Nat. Phys. {\bf 8}, 795 (2012).

\bibitem{Goldhaber} J. R. Williams, A. J. Bestwick, P. Gallagher, S. S.
Hong, Y. Cui, A. S. Bleich, J. G. Analytis, I. R. Fisher,
and D. Goldhaber-Gordon, Phys. Rev. Lett. {\bf 109}, 056803
(2012).

\bibitem{marcus_MF} H. O. H. Churchill, V. Fatemi, K. Grove-Rasmussen, M.
Deng, P. Caroff, H. Q. Xu, and C. M. Marcus, Phys. Rev.
B {\bf 87}, 241401(R) (2013).

%-Parafermions----------------------------



%ÒDisorder variables and
%parafermions in two-dimensional statistical mechanics.Ó
\bibitem{Fradkin_PF_1980} E. Fradkin and L. P. Kadanoff,  Nucl.
Phys. B {\bf 170}, 1 (1980).

%Topological Nematic States and Non-Abelian Lattice Dislocations
\bibitem{topology_barkeshli} M. Barkeshli and X.-L. Qi,
Phys. Rev. X {\bf 2}, 031013 (2012).
%arXiv:1112.3311


%ÒParafermionic edge zero modes in Zn-invariant
%spin chains,Ó
\bibitem{Fendley_PF_2012} P. Fendley,  J. Stat. Mech. {\bf 2012}, 11020 (2012).

\bibitem{Cheng} M. Cheng, Phys. Rev. B {\bf 86}, 195126 (2012).

\bibitem{PF_Linder} N. Lindner, E. Berg, G. Refael, and A. Stern, Phys.
Rev. X {\bf 2}, 041002 (2012).

\bibitem{barkeshli_2} M. Barkeshli, C, Jian, and X.-L. Qi, Phys. Rev. B {\bf 87}, 045130.

\bibitem{Vaezi} A. Vaezi, Phys. Rev. B {\bf 87}, 035132 (2013).

\bibitem{PF_Clarke} D. Clarke, J. Alicea, and K. Shtengel, Nat. Commun.
{\bf 4}, 1348 (2013).

\bibitem{Ady_FMF} Y. Oreg, E. Sela, and A. Stern, Phys. Rev. B {\bf 89}, 115402 (2014).

\bibitem{PF_Mong} R.  Mong, D. Clarke, J. Alicea, N.  Lindner, P. Fendley, C. Nayak, Y. Oreg, A. Stern, E. Berg, K. Shtengel, and M. P. A. Fisher, Phys. Rev. X {\bf 4}, 011036 (2014).

\bibitem{vaezi_2} A. Vaezi, arXiv:1307.8069.

\bibitem{PFs_Loss} J. Klinovaja and D. Loss, arXiv:1311.3259.

\bibitem{PFs_Loss_2} J. Klinovaja and D. Loss, arXiv:1312.1998.

\bibitem{PF_TI_Amir} J. Klinovaja, A. Yacoby, and D. Loss, arXiv:1403.4125.

\bibitem{Vaezi_Barkeshli} A. Vaezi and  M. Barkeshli, arXiv:1403.3383.

%%%%%


\bibitem{giamarchi_book} T. Giamarchi, {\it Quantum Physics in One Dimension} (Oxford University Press, New York, 2003).

\bibitem{QHE_graphene} Y. Zhang, Y.-W. Tan, H. L. Stormer and  P. Kim, Nature {\bf 438}, 201 (2005).


\bibitem{braunecker_prb_10} B. Braunecker, G. I. Japaridze, J. Klinovaja, and D. Loss, Phys. Rev. B \textbf{82}, 045127 (2010).

\bibitem{RSOI_1} L. Levitov and E. Rashba, Phys. Rev. B {\bf 67}, 115324 (2003).

\bibitem{RSOI_2} S. Gangadharaiah, S. Sun, and O. Starykh, Phys. Rev. Lett. {\bf 100}, 156402 (2008).

 
\bibitem{BDL} S. Bravyi, D. P. DiVincenzo, and D. Loss, Annals of Physics {\bf 326}, 2793 (2011).


\bibitem{Haldane}  F. D. M. Haldane, Phys. Rev. Lett. {\bf 61}, 2015 (1988). 






\end{thebibliography}
\end{document}